\documentclass[aps,superscriptaddress, showpacs,preprintnumbers, superscriptaddress, nofootinbibt,twocolumn]{revtex4}
\usepackage{eurosym}
\usepackage{amsfonts}
\usepackage{amssymb,amsmath,enumitem}
\usepackage{graphicx,color,epstopdf}
\usepackage{subfigure}

\def\be{\begin{equation}}
\def\ee{\end{equation}}
\def\beg{\begin{align}}
\def\eeg{\end{align}}
\def\bea{\begin{eqnarray}}
\def\eea{\end{eqnarray}}

\newcommand{\f}[2]{\frac{#1}{#2}}
\newcommand{\te}[1]{\textcolor{black}{#1}}

\begin{document}

\title{Non-minimal geometry-matter couplings in Weyl-Cartan space-times: $f(R,T,Q,T_m)$ gravity }
\author{Tiberiu Harko}
\email{tiberiu.harko@aira.astro.ro}
\affiliation{Department of Theoretical Physics, National Institute of Physics
and Nuclear Engineering (IFIN-HH), Bucharest, 077125 Romania,}
\affiliation{Astronomical Observatory, 19 Ciresilor Street, 400487 Cluj-Napoca, Romania,}
\affiliation{Department of Physics, Babes-Bolyai University, Kogalniceanu Street,
400084 Cluj-Napoca, Romania,}
\affiliation{School of Physics, Sun Yat-Sen University, Xingang Road, 510275 Guangzhou,
P. R. China,}
\author{Nurgissa  Myrzakulov}
\email{nmyrzakulov@gmai.com}
\affiliation{ Eurasian International Centre for Theoretical Physics, Nur-Sultan 010009, Kazakhstan,}
\affiliation{Eurasian  National University, Nur-Sultan 010008, Kazakhstan,}
\author{Ratbay Myrzakulov}
\email{rmyrzakulov@gmai.com}
\affiliation{ Eurasian International Centre for Theoretical Physics, Nur-Sultan 010009, Kazakhstan,}
\affiliation{Eurasian  National University, Nur-Sultan 010008, Kazakhstan,}
\author{Shahab Shahidi}
\email{s.shahidi@du.ac.ir}
\affiliation{School of Physics, Damghan University, Damghan, Iran.}

\date{\today }

\begin{abstract}
We consider an extension of standard General Relativity in which the Hilbert-Einstein action is replaced by an arbitrary function of the Ricci scalar, nonmetricity, torsion, and the trace of the matter energy-momentum tensor. By construction, the action involves a non-minimal coupling between matter and geometry. The field equations of the model are obtained, and they lead to the nonconservation of the matter energy-momentum tensor. A thermodynamic interpretation of the nonconservation of the energy-momentum tensor is also developed in the framework of the thermodynamics of the irreversible processes in open systems. The Newtonian limit of the theory is considered, and the generalized Poisson equation is obtained in the low velocity and weak fields limits. The nonmetricity, the Weyl vector, and the matter couplings generate an effective gravitational coupling in the Poisson equation.  We investigate the cosmological implications of the theory for two different choices of the gravitational action, corresponding to an additive and a multiplicative algebraic structure of the function $f$, respectively. We obtain the generalized Friedmann equations, and we compare the theoretical predictions with the observational data. \te{We find that the cosmological models can give a good descriptions of the observations up to a redshift of $z=2$, and, for some cases, up to a redshift of $z=3$ .}
\end{abstract}
\maketitle
\tableofcontents
%\newpage
\section{Introduction}

Since its proposal by Einstein \cite{Eina} and Hilbert \cite{Hil} in 1915, the development of general relativity, a geometric theory of gravitation, and, more generally, of theoretical physics, was closely intertwined with the advances in the fields of the mathematical sciences. General relativity itself is essentially based on Riemannian geometry, first presented in a consistent and systematic way in 1854 \cite{Riem}. The theory of tensors, initiated by Ricci and Levi-Civita \cite{Ric} proved to be an essential tool in the mathematical formulation of general relativity. The geometric as well as physical properties of the gravitational field are described in general relativity by the Riemann tensor $R_{\mu \nu \lambda}^{\sigma}$, and by its contractions (the Ricci tensor and scalar, respectively), with the help of which the Einstein gravitational field equations are constructed.

After its remarkable success in the field of physics, general relativity had a deep influence on mathematics, leading to several important developments. In general relativity the properties of the space-time are described by the metric tensor $g_{\mu\nu}$, satisfying the fundamental mathematical  property $\nabla _{\lambda}g_{\mu \nu}=0$, where $\nabla _{\lambda}$ is the covariant derivative constructed with the use of the Levi-Civita connection $\breve{\Gamma}_{\mu \nu}^{\lambda}$.  in 1918 Weyl, inspired by Einstein's theory,  introduced the first generalization of Riemannian geometry \cite{Weyl}, in which the covariant derivative of the metric tensor is assumed to be nonzero, $\nabla _{\lambda}g_{\mu \nu}=Q_{\lambda \mu \nu}$, which allowed to introduce a new geometric quantity, the nonmetricity $Q_{\lambda \mu \nu}$. Based on this geometry, Weyl was able to construct the first unified geometric theory of gravity and electromagnetism.  A new geometric concept, the torsion tensor $T_{\mu \nu}^{\lambda}=\Gamma _{\;\mu \nu}^{\lambda}-\Gamma_{\;\nu \mu}^{\lambda}$, where $\Gamma _{\;\mu \nu}^{\lambda}$  are the connection coefficients defining the connection, was introduced by Cartan \cite{Car1,Car2,Car3,Car4}. Finally, in this brief review of the development of the physically related mathematical ideas one must also mention the Weitzenb\"{o}ck spaces \cite{Weit}. A Weitzenb\"{o}ck space is described
by the geometric properties $\nabla_{\lambda}g_{\mu \nu} = 0$, $T_{\mu \nu}^{\lambda}\neq 0$, and $R_{\mu \nu \lambda}^{\sigma}=0$, respectively.

\te{Despite its outstanding accomplishment at the level of Solar System, general relativity was extended mathematically in several directions. Theories based on torsion were developed in the form of the Einstein-Cartan theory \cite{Hehl}, with the hope that the inclusion of torsion may solve the singularity problem that plagues standard general relativity.  Einstein was the first to use Weitzenb\"{o}ck type geometries for the
 unification of electromagnetism and  gravitation within a teleparallel theory \cite{TE1}.
In the teleparallel formulation of gravity the basic geometric quantity is considered the torsion tensor, generated by the tetrad fields,
and  with the curvature replaced by the torsion. This theoretical approach to gravity is called
the teleparallel equivalent of General
Relativity (TEGR), and it was introduced initially in \cite{TE2,TE3,TE4}. Presently, this formulation is also known as the $f(T)$ gravity
theory, with $T$ denoting the torsion scalar. The most important feature
of the $f(T)$ type theories {\it is that torsion exactly cancels curvature}, and thus
{\it the curved space-time of general relativity becomes a flat manifold}. In $f(T)$ theory instead of the metric the
tetrad fields are used as the dynamical variable, and the Lagrangian is constructed from $T$ only. In $f(T)$ theory the corrections corresponding to higher energy scales correspond to higher order terms in torsion. $f(T )$ theory has many important applications in cosmology, in the explanation of inflation \cite{TE4a}, and of the late-time cosmic acceleration \cite{TE4b}. From the point of view of the astrophysical applications, in the $f(T )$ theory a new exact charged black hole solution does exist, which contains, in addition to the monopole term, a quadrupole term, originating from the quadratic form of $f(T )$, with $f(T)\propto T^2$ \cite{TE4c}. Another major advantage of $f(T)$ gravity theory is that the
gravitational properties of the space-time are described by second order differential equations. For a detailed description of teleparallel
theories see \cite{TE5}.}

Mainly due to the criticisms by Einstein and Pauli, the Weyl geometry was mostly ignored by physicists in its first 50 years of existence.  Einstein argued that Weyl's unified theory implies the existence of the second clock
effect \cite{Str}, which implies that in Weyl geometry and in the
presence of electromagnetic fields sharp spectral lines  cannot exist, due to the dependence of the atomic clocks on their past history \cite{Str}.
According to Einstein, this predicted effect is an immediate consequence of Weyl’s theory, since the length of a vector is not constant during parallel transport. Hence, in Weyl geometry, the periods of the atomic clocks, determined by periodical physical process, must be path dependent \cite{Str}.
However,  the position of Weyl geometry changed drastically after around 1970, with more and more investigations of its stimulating physical and mathematical applications performed in both elementary particle physics and gravitational theories. Since all the equations of motion of the fields of the Standard Model of elementary particle physics (with the exception of the Higgs fields), can be derived from conformally invariant Lagrangians, the applications of Weyl geometry to quantum phenomena looks promising. Weyl geometry offers the possibility of including conformal, or scale covariant classical relativity into the standard model of elementary particles, and to explore the relation between classical relativity and quantized theories of gravity (for a review of the applications of Weyl geometry in physics see \cite{Scholz}.

 An important generalization of Weyl gravity was introduced by Dirac \cite{Dirac1,Dirac2}. By introducing a real scalar field $\beta$ of weight $w(\beta)=-1$,  Dirac proposed for the gravitational Lagrangian the expression
\te{ \be\label{LD}
 L=-\beta ^2R+kD^{\mu}\beta D_{\mu}\beta +c\beta ^4+\frac{1}{4}W_{\mu \nu}W^{\mu \nu},
 \ee}
 where the electromagnetic type field tensor \te{$W_{\mu \nu}$ is constructed from the Weyl length curvature} and the Weyl connection vector $w_{\mu}$ according to the definition $W_{\mu \nu}=\nabla _{\nu}w_{\mu}-\nabla_{\mu}w_{\nu}$, and $k=6$ is a constant. The Lagrangian given by Eq.~(\ref{LD} is conformally invariant. The cosmological applications of a modified Dirac Lagrangian were considered in \cite{Rosen}. Another Weyl-Dirac type Lagrangian was proposed in \cite{Isrcosm}, and it is given by
 \bea
 L&=&W^{\lambda \rho}W_{\lambda \rho}-\beta ^2R+\sigma \beta ^2w^{\lambda}w_{\lambda}+2\sigma \beta w^{\lambda}\beta _{,\lambda}+\nonumber\\
&& (\sigma +6)\beta _{,\rho}\beta_{,\lambda }g^{\rho \lambda}+2\Lambda \beta ^4+L_m,
 \eea
 where $\beta $ denotes again the Dirac scalar field, while $\sigma$ and $\Lambda$ are constants. The cosmological evolution of a Universe described by the above Lagrangian was also considered.  In this model ordinary matter is generated at the beginning of the Universe due to the presence of Dirac’s gauge function. Interestingly enough, in the late Universe, filled with pressureless matter, Dirac’s gauge function generates the dark energy that causes the de Sitter type cosmic acceleration.

 Conformal Weyl gravity, quadratic in the scalar curvature, and in the Weyl tensor, was investigated, in both metric and Palatini formulations, in \cite{Gh1,Gh2,Gh3,Gh4,Gh5,Gh6,Gh7}. The elementary particle physics as well as its implications for the very early Universe evolution were investigated. \te{The quadratic Weyl action has spontaneous symmetry breaking in a Stueckelberg mechanism, with the result that the Weyl gauge field becomes massive.} Hence, one recovers the Einstein-Hilbert action of standard general relativity in the presence of a positive cosmological constant, together with the Proca action for the massive Weyl gauge field. The action is \cite{Gh7}
\bea
L_0=\sqrt{-g} \,\Big[\, \frac{1}{4!}\,\frac{1}{\xi^2}\,\tilde R^2  - \frac14\, F_{\mu\nu}^{\,2}
-\frac{1}{\eta^2}\,\tilde C_{\mu\nu\rho\sigma}^{\,2}\Big],
\eea
where $\xi, \eta\leq 1$ are coupling constants,  $F_{\mu\nu}$ is the field strength of the Weyl vector $w_\mu$,
 $\tilde R$ is the scalar curvature in Weyl geometry, while $\tilde C_{\mu\nu\rho\sigma}$
is the  Weyl tensor. By replacing  $\tilde R^2$ with $ -2\phi_0^2 \tilde R-\phi_0^4$  \cite{Gh7}, where $\phi_0$ a scalar field, and after a rescaling of the variables one obtains the action
\bea
L_0&=&\sqrt{-g} \,\Big[- \frac{ M_p^2}{2}\hat R +\frac{3}{4} M_p^2\alpha ^2\,\gamma^2\hat\omega_\mu \hat \omega^\mu - \frac{3}{2} \xi^2 M_p^4 \nonumber\\
&&-\frac{1}{4} \hat F_{\mu\nu}^2-\frac{1}{\eta^2}  C_{\mu\nu\rho\sigma}^2\Big],
\eea
where $M_p^2= \langle\phi_0^2\rangle/6\,\xi^2$, is the Planck mass, and with the Weyl vector field $\hat\omega ^\mu$ satisfying the constraint  $\nabla_\mu\hat\omega ^\mu=0$. Hence by the Stueckelberg mechanism $\omega_\mu$ generates a massive Proca field. Moreover, in this model the number of degrees of freedom is conserved, and the massless scalar field $\phi_0$  becomes massive, and the massless field $\omega _\mu$ is substituted by a massive field $\omega _\mu$ with a mass $m_\omega ^2=(3/2) \alpha ^2\gamma^2 M_p^2$ \cite{Gh7}.

An interesting application of Weyl geometry is the so-called $f(Q)$ gravity theory, also known as symmetric teleparallel gravity. This theory was proposed in \cite{Q1}, and it represents a geometric approach to gravitation in which the nonmetricity $Q$ of a Weyl geometry becomes the basic variable describing all the physical properties of the gravitational interaction. This pathway to gravity was later developed into the  $f(Q)$ gravity theory, also known as nonmetric gravity \cite{Q2}. Different geometrical, physical and cosmological aspects of the $f(Q)$ theory have been investigated in \cite{Q3,Q4,Q5,Q6,Q7,Q8,Q9,Q10,Q11,Q12,Q13,Q14,Q15,Q16,Q17, Q18, Q19}.

 Weyl's geometry can be extended naturally to include torsion. The corresponding geometry is called the Weyl-Cartan geometry, and it was extensively studied from both physical and mathematical points of view \cite{WC1,WC2,WC3,WC4,WC5,WC6,WC7,WC8, WC9}. For a review of the geometric properties and of the physical applications and of the Riemann-Cartan and Weyl-Cartan space-times  see \cite{Rev}.

 The surprising discovery of the late acceleration of the Universe \cite%
{1n,2n,3n,4n,acc}, implying a transition from deceleration to acceleration at small redshifts $z$ of the order of $z\approx 0.6$ raised serious questions about the theoretical foundations of the most successful gravitational theory presently known, Einstein's general relativity. The simplest possibility of giving a reason to the late de Sitter type phase is to reintroduce again in the field equations the cosmological constant $\Lambda$, postulated by Einstein in 1917 in order to build the first, static general relativistic cosmological model \cite{Ein}.  The $\Lambda$ extension of the field equations is the theoretical basis of the standard cosmological paradigm, the $\Lambda$CDM model. The $\Lambda$CDM model also requires the presence of another mysterious (and still undetected) component of the Universe, the dark matter \cite{Sal}. The $\Lambda$CDM model fits very well the cosmological observations \cite{C1,C2,C3, C4}. However, it faces a major theoretical problem:  no basic physical theory can explain it. The main complication originates from the troubles met when trying to explain the origin and nature of $\Lambda$ \cite{Wein1,Wein2, Wein3}. A possible interpretation of $\Lambda$ as the Planck-scale vacuum energy density $\rho _{vac}$ leads to the ``worst prediction in physics" \cite{Lake}, since
\bea
\rho _{vac}&\approx& \frac{\hbar}{c}\int _{k_{dS}}^{k_{Pl}}{\sqrt{k^2+\left(\frac{mc}{\hbar}\right)^2}d^3k}\nonumber\\
&\approx & \rho _{Pl}=\frac{c^5}{\hbar G^2}=10^{93}\; {\rm \frac{g}{cm^3}},
\eea
a result that disagree by a factor of around $10^{-120}$  from the observed value of the energy density associated to $\Lambda$, $\rho _{\Lambda}=\Lambda c^2/8\pi G\approx 10^{-30}\;{\rm g/cm^3}$ \cite{C3}.

The $\Lambda $CDM model faces, even at the observational level, some (yet unsolved) problems.  The most important is the ``Hubble tension", which originates from the serious differences between the values of the Hubble constant, $H_0$, as obtained from the CMB measurement \cite{C4}, and the values obtained directly from observations in the local Universe \cite{M1,M2,M3}. The SH0ES determination of $H_0$ gives the value
$H_0 = 74.03 \pm 1.42$ km/s/Mpc \cite{M1}. On the other hand, from the early Universe
surveys performed by the Planck satellite one obtains  $H_0 = 67.4 \pm 0.5$ km/s/Mpc \cite{C3}, a value that
differs by $\sim  5\sigma$ from the SH0ES result.

Moreover,  many essential theoretical questions cannot be explained within the $\Lambda$CDM paradigm, like, for example, the smallness of $\Lambda$, its fine tuning, and why the transition from deceleration to acceleration took place only recently.  And, the fundamental question is if a cosmological constant is really necessary for both observational and theoretical cosmology?
Hence, the investigation of alternative avenues for the description of the gravitational interaction may give us the possibility of solving the cosmological problems without resorting to a cosmological constant. Actually, there are (at least) three theoretical possibilities that could replace  the $\Lambda$CDM paradigm.

The first possibility  is called the {\it dark components model}, and it generalizes the Einstein field equations {\it by adding two new terms in the total energy momentum tensor of the Universe}. These two new "matter" terms correspond to dark energy and dark matter, respectively. Hence, the gravitational phenomena are described in this approach by the field equation \cite{HL20}
\be
G_{\mu \nu}=\kappa ^2 T^{\rm bar}_{\mu \nu}+\kappa ^2T^{\rm DM}_{\mu \nu}(\phi, \psi _{\mu},...)+\kappa ^2T^{\rm DE}_{\mu \nu}(\phi, \psi _{\mu},...),
\ee
where $G_{\mu \nu}$ is the Einstein tensor, while $T^{\rm bar}_{\mu \nu}$, $T^{\rm DM}_{\mu \nu}(\phi, \psi_{\mu},...)$, and $T^{\rm DE}_{\mu \nu}(\phi, \psi_{\mu},...)$ represent the energy-momentum tensors of the baryonic matter, dark matter and dark energy, respectively.  The energy-momentum tensors of dark matter and dark energy are constructed from some scalar $\phi$ or vector fields $\psi _{\mu}$.  The simplest dark components model  assumes that dark energy can be described by a scalar field $\phi$, having a self-interaction potential $V(\phi)$. Thus, the gravitational action takes the form
\be
S=\int{\left[\frac{M_{p}^2}{2}R-\left(\partial \phi\right)^2-V(\phi)\right]\sqrt{-g}d^4x}.
\ee
The corresponding dark energy models are called quintessence models \cite{Qa1,Qa2,Qa3,Qa4,Qa5, Qa6}.  Many other dark component models have been proposed, like, for example,  k-essence models \cite{K1,K2,K3}, tachyon  \cite{T1,T2}, phantom \cite{Ph1,Ph2,Ph3},  quintom \cite{Qu1,Qu2,Qu3} and chameleon \cite{Ch1,Ch2,Ch3,nonlocal,Ch4} field models, as well as Chaplygin gas \cite{Cha1,Cha2}, and vector field \cite{V1,V2,V3} dark energy  models, respectively. For reviews of the dark energy models see \cite{Rev1,Rev2,Rev3,Rev4}.
Despite their remarkable success, the dark component models still face their own intrinsic problems. For example, in  \cite{Qa6} it was found that quintessence always lowers $H_0$ with respect to the $\Lambda$CDM model, with the Hubble tension becoming worse.

The second possible generalized approach to the gravitational interaction is the {\it dark gravity} formalism, which, similarly to Einsteinian relativity, is based on an entirely geometrical description of gravity. In this approach the dynamics and evolution of the Universe is explained by the modification of geometry of the space-time. In the dark gravity model, the Einstein equations are written down as
\be
G_{\mu \nu}=\kappa ^2T_{\mu \nu}^{(mat)}+\kappa ^2 T_{\mu \nu}^{(\rm geom)}\left(g_{\mu \nu}, R, Q, T, \square R,...\right),
\ee
 where $T_{\mu \nu}^{(mat)}$ is the matter energy-momentum tensor, defined in the usual way,  while $T_{\mu \nu}^{(\rm geom)}\left(g_{\mu \nu}, R, Q,T, \square R,...\right)$, which induces an effective energy-momentum tensor, which can mimic dark energy and dark matter, respectively, is a purely geometric quantity, built up from the metric, Ricci scalar, nonmetricity and torsion, respectively.  $T_{\mu \nu}^{(\rm geom)}\left(g_{\mu \nu}, R, Q, T, \square R,...\right)$ can describe the gravitational dynamics. Dark gravity approaches were first introduced via the $f(R)$ theory, proposed initially in \cite{Bu1}, and later studied in \cite{Bu2,Bu3,Bu4,Bu5}. In $f(R)$ theory the Hilbert-Einstein action $S=\int{\left(R/\kappa ^2+L_m\right)\sqrt{-g}d^4x}$ of general relativity is changed by the action  $S= \int{\left[f(R)/\kappa^2+L_m\right]\sqrt{-g}d^4x}$, where $f(R)$ is an analytical function of the Ricci scalar $R$. The cosmological and astrophysical implications of the $f(R)$ model have been studied intensively  \cite{fR1,fRn1,fRn2,fRn3,fR2,fR3,fR4,fR5,fR6,fRn4, fR7,fR8,fR9,fR10,fR11}. In \cite{fRn4} the first internally consistent $f(R)$ model was developed, with the model describing both early inflation, as well as late dark energy. The Hybrid Metric-Palatini Gravity theory \cite{HMP1,HMP2,HMP3,HMP4} represents another purely geometric theory that generalizes and unifies two different geometric approaches, the metric and the Palatini ones, respectively. Geometric structures extend the Riemannian one, like, for example, Weyl geometry, have also been studied \cite{W1,W2,W3,W4,W5,W6,W7,W8,W9,W10,weylcartan}. For recent reviews of dark gravity theories, and their applications, see \cite{R1,R2,R3,R4, R5}.

Another avenue for understanding gravitational phenomena is represented by {\it the dark coupling} approach.  Here the basic idea is the replacement of the Hilbert-Einstein gravitational Lagrangian, which has a simple additive structure in the curvature and matter terms,  with a more general function. Hence, one can consider {\it a maximal extension of the Hilbert-Einstein Lagrangian} by assuming that the gravitational action can be represented by an arbitrary analytical function of the curvature scalar $R$, of the nonmetricity $Q$, torsion $T$, of the matter Lagrangian $L_m$, of the trace $T_m$ of the energy-momentum tensor, and, perhaps of other thermodynamic quantities. The dark coupling approach naturally determines the presence of a {\it nonminimal coupling between geometry and matter}.

In the dark coupling approach,  the Einstein gravitational field equations are given by
\bea
G_{\mu \nu}&=&\kappa ^2T_{\mu \nu}^{(mat)}\nonumber\\
&&+\kappa ^2 T_{\mu \nu}^{(\rm coup)}\left(g_{\mu \nu},  R, Q,T,L_m, T_m, \square R, \square T,... \right).\nonumber\\
\eea

The effective energy-momentum tensor of the dark coupling theories $T_{\mu \nu}^{(\rm coup)}\left(g_{\mu \nu}, R, Q, T,L_m, T_m, \square R, \square T,... \right)$ is obtained from a {\it non-additive geometry-matter algebraic structure}, which involves the couplings between all forms of matter, and all forms of scalar geometric quantities.

The simplest form of dark coupling type theories was proposed initially in \cite{fLm1}, where the gravitational action
\be
S=\int{\left[f_1(R)+\left(1+\lambda f_2(R)\right)L_m\right]\sqrt{-g}d^4x},
\ee
was studied. This action was generalized in \cite{fLm2}, and in  \cite{fLm3}, thus leading to the $f\left(R,L_m\right)$ gravity theory, with the action given by $S=\int{f\left(R,L_m\right)\sqrt{-g}d^4x}$. Various astrophysical and cosmological applications of the $f\left(R,L_m\right)$ theory, as well as its fundamental aspects were investigated in \cite{fLm4, fLm5,fLm6,fLm6a, fLm6b, fLm7,fLm8,fLm9,fLm10,fLm11, fLm12}.

 A different possibility of coupling between geometry and matter is considered in the $f(R,T_m)$ gravity theory \cite{fT1}, in which geometry, represented by the Ricci scalar $R$, is nonminimally coupled to the trace of the matter energy-momentum tensor $T_m$.  The action of the theory is given by $S=\int{\left[f\left(R,T\right)+L_m\right]\sqrt{-g}d^4x}$. The astrophysical and cosmological applications of $f(R,T)$ theory were studied extensively in \cite{fT2,fT3,fT4,fT5,fT6,fT7,fT8,fT9,fT10,fT11,fT12,fT12a, fT13,fT13a, fT14, fT14a, fT14b, fT15, fT15a, fT16, fT17, fT18, fT19, fT20, fT21, fT22}. A generalization  of the $f(R,T)$ gravity theory was proposed in \cite{fT23}, by introducing higher derivatives matter fields, with the action given by
 \be
 S=\frac{1}{16\pi}\int{f\left(R,T_m,\Box T_m\right)\sqrt{-g}d^4x}+\epsilon \int{L_m\sqrt{-g}d^4x},
 \ee
 where $\epsilon =\pm 1$. Accelerated expansion of the de Siter type naturally emerges in this scenario in the pressureless matter filled Universe, without the need of any additional matter or geometric components. The model can also describe inflationary evolution, with results in good agreement with observations. An extension of $f(T)$ gravity, allowing for a general coupling of the torsion scalar $T$ with the trace of the matter energy-momentum tensor $T_m$ was introduced in \cite{fTT}. The resulting $f\left(T,T_m\right)$ theory is a new modified gravity, since it is different from all the existing torsion or curvature based constructions. As applied to a cosmological framework, it leads to interesting phenomenology, and can be considered as a promising candidate to explain the accelerated expansion of the Universe. The coupling of geometry and matter in $f(Q)$ gravity was considered in \cite{fQC1,fQC2,fQC3}.

 It is the main goal of the present paper to consider {\it the maximal extension of the Hilbert-Einstein variational principle on a metric-affine manifold, endowed with a connection containing the Levi-Civita connection, as well as the contorsion and the disformation tensors}. In the presence of matter on this manifold we consider four scalar quantities $\left(R,Q,T,T_m\right)$, from which three $\left(R,Q,T\right)$ have a purely geometric origin, while $T_m$ describes the matter properties. The maximal extension of the Hilbert-Einstein variational principle is then represented by an arbitrary analytical function of the four scalars, $L_g=f\left(R,Q,T,T_m\right)$. We assume that $f$, as an  analytic function can be locally given by a convergent power series. Moreover, the Taylor series of $f$ about $x_0=\left(R_0,Q_0,T_0,T_m^{(0)}\right)$ converges to the function $f$ in some neighborhood for every $x_0$ in its domain.

  By definition, the action constructed from the gravitational Lagrangian density $f$ involves a non-minimal coupling between matter and geometry. As a first step in our study we derive the field equations of the model. By considering the covariant divergence of the field equations we obtain the covariant divergence of the matter energy-momentum tensor,  which generally does not vanish. Hence the present theory leads to the nonconservation of the matter energy-momentum tensor. A possible thermodynamic interpretation of the nonconservation of the energy-momentum tensor is also briefly discussed in the framework of the thermodynamics of the irreversible processes in open systems.  We also consider the Newtonian limit of the theory in the low velocity and weak gravitational field limits, and the generalized Poisson equation is obtained, involving a varying effective gravitational constant.  The cosmological implications of the theory for different choices of the gravitational action are also investigated in detail. More exactly, we consider additive and multiplicative algebraic structures for the gravitational Lagrangian.  For each case we obtain the generalized Friedmann equations, and we compare the theoretical predictions with the observational data. We find that the models can give a good descriptions of the observations at least up to a redshift of $z=2$, for some models, even up to a redshift of $z\approx 3$.

The present paper is organized as follows. The action $S$ of the theory is written down in Section \ref{sect1}, and the field equations are derived by varying $S$ with respect to the metric tensor. The representations of the nonmetricity and of the torsion tensor are also introduced. The divergence of the matter energy-momentum tensor, the Newtonian limit of the theory, leading to the generalized Poisson equation, and the thermodynamic interpretation of the model are considered in Section~\ref{sect2}. The cosmological implications of the theory are considered in Section~\ref{sect3}, where two classes of models, corresponding to an additive and multiplicative structure of the gravitational Lagrangian are considered. The generalized Friedmann equations are obtained, and the predictions of the theory are compared with the $\Lambda$CDM model, and with the observational data. Finally, we discuss and conclude our results in Section~\ref{sect4}.

%%%%%%%%%%%%%%%%%%%%%%%%%%%%%
\section{Action and gravitational field equations}\label{sect1}
%%%%%%%%%%%%%%%%%%%%%%%%%%%%%

We consider that the gravitational action, defined in a Weyl-Cartan geometry, can be constructed in the following way \te{\cite{M1,M1a,M1b,M1c}},
\begin{equation}\label{2.1}
S(g,\Gamma, \phi)=S_{g}+S_{m}=\int \sqrt{-g}d^{4}x \left[f\left(R,T,Q,T_m\right)+L_{m}\right],
\end{equation}
where $f$ is an arbitrary analytical function (it can be represented locally by convergent power series expansion), $R$ stands for the Ricci scalar (curvature scalar), $T$ is the torsion scalar, $Q$ is the nonmetricity scalar and,  $T_m$ is the trace of the energy-momentum tensor of the matter $T_{\mu \nu}$, obtained from the Lagrangian $L_{m}$. The theory derived from the above Lagrangian can be considered as the unification of the $f(R)$ \cite{Bu1}, $f(T)$ \cite{TE5}, $f(Q)$ \cite{Q2},  $f\left(R,T_m\right)$ \cite{fT1}, $f\left(T,T_m\right)$ \cite{fTT},  and $f\left(Q,T_m\right)$ \cite{fQC1, fQC2} type theories, respectively. Theoretical gravitational models with action of the form by $f(R,T,Q,T_m,D)$ where $D$ is the divergence of the dilation current have also been investigated \cite{M1a}.

Since our model, given by the action Eq.~(\ref{2.1}), lives in a metric-affine manifold $\left(M, g_{\mu\nu}, \Gamma^{\rho}_{\,\,\mu\nu}\right)$, let us briefly present some basic differential geometric properties of this manifold. The corresponding  connection $\Gamma^{\rho}_{\,\,\mu\nu}$ decomposes as
\begin{equation}
\Gamma^{\rho}_{\,\,\mu\nu}=\breve{\Gamma}_{\,\, \mu \nu}^{\rho}+K^{\rho}_{\,\,\mu\nu}+L^{\rho}_{\,\,\mu\nu}\,,
\end{equation}
where $\breve{\Gamma}_{\,\, \mu \nu}^{\rho}$ is the Levi--Civita connection,   $K^{\rho}_{\,\,\mu\nu}$  is  the  contorsion tensor and
$L^{\rho}_{\,\,\mu\nu}$ is  the disformation tensor.

These  tensors  have the following forms
\begin{align}
\breve{\Gamma}^\alpha_{~\beta\gamma}=& \frac{1}{2} g^{\alpha\rho} \left( \partial _\gamma g_{\beta\rho} + \partial _\beta g_{\gamma\rho} - \partial _\rho g_{\beta\gamma} \right), \\
K_{\mu\nu\alpha}=&\frac{1}{2}\bigl(T_{\mu\nu\alpha}+T_{\nu\alpha\mu}-T_{\alpha\mu\nu}\bigr),\\
S_\alpha^{~\mu\nu}=&\frac12\bigl(K^{\mu\nu}_{~~~\alpha}+\delta^\mu\alpha T^{\beta\nu}_{~~~\beta}-\delta^\nu_\alpha T^{\beta\mu}_{~~\beta}\bigr),\\
L_{\alpha\mu\nu}=&\frac{1}{2}\bigl(Q_{\alpha\mu\nu}-Q_{\mu\nu\alpha}-Q_{\nu\mu\alpha}\bigr).
\end{align}

Here we have also introduced the torsion and the non-metricity tensors, defined according to
\begin{equation}
T_{\,\,\,\, \mu \nu}^{\alpha}=2 \Gamma_{\,\,\,\, [\mu \nu]}^{\alpha}\,, \quad Q_{\rho \mu \nu} = \nabla_{\rho} g_{\mu \nu}.
\end{equation}

In this metric-affine  spacetime, let us introduce three scalars as
\begin{align}
 R=&g^{\mu\nu}R_{\mu\nu},\label{1.3}\\
 T=&{S_\rho}^{\mu\nu}\,{T^\rho}_{\mu\nu},\label{1.5}\\
Q=& L^\alpha_{~\beta\alpha}L^{\beta\mu}_{~~~\mu}-L^\alpha_{~\beta\mu}L^{\beta\mu}_{~~~\alpha},
 \end{align}
 where $R$ is {\it the curvature scalar of the Weyl-Cartan geometry}, $T$ is the torsion scalar, and $Q$ is the nonmetricity  scalar, respectively. The Ricci tensor is constructed from the affine connection according to its standard definition
\begin{align}
R_{\nu\beta}&=\partial_\mu\Gamma^\mu_{~\nu\beta}-\partial_\nu\Gamma^\mu_{~\mu\beta}+\Gamma^\mu_{~\mu\alpha}\Gamma^\alpha_{~\nu\beta}-\Gamma^\mu_{~\nu\alpha}\Gamma^\alpha_{~\mu\beta}.
 \end{align}

Varying the action Eq.~(\ref{2.1})  with respect to the metric and the connection,  we obtain the  following two field equations \cite{M1a}
\begin{widetext}
\begin{align}
f_R&\Big[g_{\mu[\nu}Q_{\alpha]\beta}^{~~~~\beta}+T_{\mu\alpha\nu}+Q_{\alpha\mu\nu}-g_{\alpha\mu}Q_{\beta\nu}^{~~~\beta}\Big]+f_Q\Big[2g_{\mu(\nu}L^\beta_{~\alpha)\beta}+g_{\alpha\nu}(L_{\mu\beta}^{~~~\beta}-L^\beta_{~\mu\beta})-2L_{\mu\alpha\nu}\Big]\nonumber\\
&+f_T\Big[T_{\alpha\mu\nu}+T_{\mu\alpha\nu}-T_{\nu\alpha\mu}+g_{\alpha[\nu}T^\beta_{~~\mu]\beta}\Big]+2g_{\mu[\alpha}\nabla_{\nu]}f_R-\frac12H_{\alpha\mu\nu}=0,
	\end{align}
\end{widetext}
and
\begin{widetext}
\begin{align}
f_R R_{\mu\nu}&-\frac12f\,g_{\mu\nu}+f_{T_m}(g_{\mu\nu}L_m-T_{\mu\nu})-\frac12T_{\mu\nu}-\frac12\nabla_\alpha\Big(A^\alpha_{~(\mu\nu)}-A_{(\mu~~\nu)}^{~~\,\alpha}+A_{(\mu\nu)}^{~~~~\alpha}\Big)\nonumber\\
&+\frac14f_T\Big[2T_{\alpha\nu\beta}T^{\alpha~~\beta}_{~~\mu}+2T^{\alpha~~\beta}_{~~\mu}T_{\beta\nu}^{~~~\alpha}-4T^\alpha_{~~\mu\alpha}T^\beta_{~~\nu\beta}-T_\mu^{~~\alpha\beta}T_{\nu\alpha\beta}\Big]\nonumber\\
&+\frac14f_Q\Big[3Q_{\alpha~~\beta}^{~~\beta}Q^\alpha_{~\mu\nu}-g_{\mu\nu}Q_{\alpha\beta}^{~~~\beta}Q^{\alpha\gamma}_{~~~\gamma}-2Q_{\alpha\nu\beta}Q^{\alpha~~\beta}_{~\,\mu}+2Q^{\alpha~~\beta}_{~~\mu}Q_{\beta\nu\alpha}-2Q^\alpha_{~\mu\alpha}Q^\beta_{~\nu\beta}\nonumber\\
&+Q_{\beta(\mu}^{~~~~\beta}Q_{\nu)\alpha}^{~~~~\alpha}-2Q_{\alpha~~\beta}^{~~\beta}Q_{(\mu\nu)}^{~~~~\alpha}+Q_\mu^{~~\alpha\beta}Q_{\nu\alpha\beta}\Big]=0,
\end{align}
\end{widetext}
respectively, where  {\it the energy-momentum tensor} and {\it the hypermomentum tensor} are given by
\begin{eqnarray}
{ T}_{\mu\nu}=-\frac{2}{\sqrt{-g}}\frac{\delta S_{m}}{\delta g^{\mu\nu}}, \quad H_{\lambda}^{\, \, \, \mu\nu}=-\frac{2}{\sqrt{-g}}\frac{\delta S_{m}}{\delta \Gamma^{\lambda}_{\,\,\, \mu\nu}}.
\end{eqnarray}
\te{Also, we have defined
\begin{align}
A_{\mu\alpha\nu}=f_Q\big(g_{\mu\nu}L^\beta_{~\alpha\beta}+g_{\alpha\nu}L_{\mu\beta}^{~~\beta}-L_{\mu\alpha\nu}-L_{\nu\alpha\mu}\big).
\end{align}}
It should be noted that {\it all the curvature terms and derivatives are constructed from the affine connection $\Gamma^\alpha_{~\mu\nu}$}.

\subsection{Introducing the  Weyl vector}

In this Subsection, we will assume a simple form for the non-metricity tensor, and define it according to
\begin{align}
Q_{\mu\nu\alpha}=w_\mu g_{\nu\alpha},
\end{align}
where $w_{\mu}$ is the {\it Weyl vector}. In this case the gravitational field equations of the $f\left(R,T,Q,T_m\right)$ theory simplify to
\begin{align}
f_T&\Big[T_{\alpha\mu\nu}-g_{\alpha[\mu}T^\beta_{~~\nu]\beta}+T_{\mu\alpha\nu}-T_{\nu\alpha\mu}\Big]\nonumber\\&+f_Q\Big[2g_{\alpha\nu} w_\mu-2g_{\mu(\alpha}w_{\nu)}\Big]\nonumber\\&+f_R\Big[T_{\mu\alpha\nu}-2g_{\mu[\alpha}w_{\nu]}\Big]+2g_{\mu[\nu}\nabla_{\alpha]}f_R-\frac12H_{\alpha\mu\nu}=0,
\end{align}
and
\begin{widetext}
\begin{align}
f_R R_{\mu\nu}&-\frac12f\,g_{\mu\nu}+f_{T_m}(g_{\mu\nu}L_m-T_{\mu\nu})-\frac12T_{\mu\nu}-g_{\mu\nu}w^\alpha\nabla_\alpha f_Q+w_{(\nu}\nabla_{\mu)}f_Q\nonumber\\&
+\frac12f_Q\Big[\nabla_{(\mu}w_{\nu)}-5g_{\mu\nu}w^\alpha w_\alpha-w_\mu w_\nu-g_{\mu\nu}\nabla_\alpha w^\alpha-g_{\alpha(\mu}\nabla_{\nu)}w^\alpha\Big]
\nonumber\\&+\frac14f_T\Big[2\,T_{\alpha\nu\beta}T^{\alpha~~\beta}_{~~\mu}+2\,T^{\alpha~~\beta}_{~~\mu}T_{\beta\nu}^{~~~\alpha}-4T^\alpha_{~~\mu\alpha}T^\beta_{~~\nu\beta}-T_\mu^{~~\alpha\beta}T_{\nu\alpha\beta}\Big]=0,
\end{align}
\end{widetext}
respectively. We note at this moment that in the above equations, {\it all the curvature terms and their derivatives are constructed from the affine connection $\Gamma^\alpha_{~\mu\nu}$}.

\subsection{Decomposition of the torsion tensor}

The torsion tensor $T_{\mu\nu\alpha}$ can be irreducibly decomposed as
\begin{align}
T_{\mu\nu\rho}=\frac23(t_{\mu\nu\rho}-t_{\mu\rho\nu})+(A_\nu g_{\mu\rho}-A_\rho g_{\mu\nu})+\epsilon_{\mu\nu\rho\sigma}B^\sigma,
\end{align}
where $t_{\mu\nu\alpha}$ is a tensor, with the following properties,
\begin{align}
t_{\mu\nu\rho}+t_{\nu\rho\mu}+t_{\rho\mu\nu}=0,\quad t^\mu_{~\,\mu\nu}=0=t^\mu_{~\,\nu\mu},
\end{align}
and $3A_\mu=T^\alpha_{~\mu\alpha}$ and $B_\mu$ {\it are two arbitrary vectors}. In the present work, {\it we will assume that only the $A_\mu$ vector is non-zero}. As a result, the torsion tensor can be written as
\begin{align}
T_{\mu\nu\alpha}=A_\nu g_{\mu\alpha}-A_\alpha g_{\mu\nu}.
\end{align}

The field equations simplify in this case as
\begin{align}\label{affineq}
8&f_T g_{\alpha[\mu}A_{\nu]}+2f_Q\Big[g_{\alpha\nu}w_\mu-g_{\mu(\nu}w_{\alpha)}\Big]\nonumber\\&+2f_R\Big[g_{\mu[\nu}A_{\alpha]}-g_{\mu[\alpha}w_{\nu]}\Big]+2g_{\mu[\alpha}\nabla_{\nu]}f_R=0,
\end{align}
and
\begin{widetext}
\begin{align}\label{metriceq}
f_R \breve{R}_{\mu\nu}&-\frac12f\,g_{\mu\nu}+f_{T_m}(g_{\mu\nu}L_m-T_{\mu\nu})-\frac12T_{\mu\nu}-g_{\mu\nu}w^\alpha\breve{\nabla}_\alpha f_Q+w_{(\nu}\breve{\nabla}_{\mu)}f_Q-f_T\,A_\mu A_\nu\nonumber\\&
+\frac12f_Q\Big[3A^\alpha w_\alpha g_{\mu\nu}-2g_{\mu\nu}w_\alpha w^\alpha-3A_{(\mu}w_{\nu)}+2w_\mu w_\nu-2g_{\mu\nu}\breve{\nabla}_\alpha w^\alpha+2\breve{\nabla}_{(\mu}w_{\nu)}\Big]\nonumber\\&
+\frac12 f_R\Big[2(A_\alpha A^\alpha g_{\mu\nu}-A_\mu A_\nu)+A^\alpha w_\alpha g_{\mu\nu}+2+A_{(\mu}w_{\nu)}+w_\mu w_\nu-w^\alpha w_\alpha g_{\mu\nu}\nonumber\\&-g_{\mu\nu}\breve{\nabla}_\alpha(2A^\alpha-w^\alpha)+\breve{\nabla}_{(\mu}A_{\nu)}+\breve{\nabla}_{(\mu}w_{\nu)}\Big]=0,
\end{align}
\end{widetext}
respectively. In the above equations, {\it all the curvature terms and the derivatives are constructed from the Levi-Civita connection $\breve{\Gamma}^\alpha_{~\mu\nu}$}. On the other hand {\it we have assumed that the energy-momentum tensor of the  matter field is constructed only from the metric tensor, and it is independent from the affine connection}.

For later convenience, let us recall that the torsion, non-metricity and curvature scalars can be written as
\begin{align}
T&=-6A^2,\quad Q=-\frac32 w^2,\nonumber\\
R&=\breve{R}+3A^2+3A_\alpha w^\alpha-\frac32w^2-3\breve{\nabla}_\mu A^\mu+3\breve{\nabla}_\mu w^\mu,
\end{align}
with $A^2\equiv A_\alpha A^\alpha$ and $w^2\equiv w_\alpha w^\alpha$.

Let us note about a special case where the torsion vector can be obtained from the Weyl vector as
\begin{align}
A_\mu=\alpha w_\mu.
\end{align}

In this case, from the connection equation \eqref{affineq}, one obtains $f_T=1/2\alpha f_Q$, which implies that the torsion tensor is not independent. Also, we obtain,
\begin{align}
\Big[f_Q+(1-\alpha)f_R\Big]w_\alpha+\nabla_\alpha f_R=0.
\end{align}
Using the above relations, one can find the metric field equation as
\begin{align}
f_R &\breve{R}_{\mu\nu}-\frac12f\,g_{\mu\nu}+f_{T_m}(g_{\mu\nu}L_m-T_{\mu\nu})\nonumber\\&+g_{\mu\nu}\Box f_R-\nabla_\mu\nabla_\nu f_R+\left(\alpha-\f12+\f3\alpha\right)\nabla_{(\mu}\big(w_{\nu)}f_R\big)\nonumber\\&+f_R\Bigg[\f12(1+\alpha)w^2g_{\mu\nu}-\left(2+\f\alpha2-\f3\alpha\right)w_\mu w_\nu\nonumber\\&+3\left(\f12-\f1\alpha\right)\nabla_{(\mu}w_{\nu)}\Bigg]=-\frac12T_{\mu\nu}.
\end{align}
\te{Let us talk about the dynamical degrees of freedom of the theory. The original theory has two independent fields, i.e. the metric tensor which is symmetric and the non-symmetric affine connection, which has 74 independent components. In field equations \eqref{affineq} and \eqref{metriceq}, we have defined the Weyl vector and the torsion vector which reduces the dof of the theory to 18. However, equation \eqref{affineq} gives an algebraic equation determining the Weyl and torsion tensors in terms of the metric. At last, the theory is left with the usual metric dof as in f(R) gravity.}
\section{Divergence of the matter energy-momentum tensor, the Newtonian limit, and thermodynamic interpretation}\label{sect2}

Taking the covariant derivative of the metric field equation and using the connection equation \eqref{affineq}, one obtains the conservation equation of the energy-momentum tensor as
\begin{widetext}
\begin{align}\label{cons}
\nabla_\alpha T^\alpha_{~\nu}&=\f{1}{2(1+2f_{T_m})}\Bigg[\nabla_\alpha\Big((6w^\alpha w_\nu-3A^\alpha w_\nu-17A_\nu w^\alpha)f_Q\Big)+2\nabla_\nu\Big((4A^\alpha w^\alpha-3w^2)f_Q\Big)\nonumber\\&+2f_Q\Big(3w_\nu\nabla_\alpha A^\alpha-4w_\nu\nabla_\alpha w^\alpha-2w^\alpha\nabla_\alpha A_\nu+5w^\alpha \nabla_\nu A_\alpha+\nabla_\nu w^2\Big)+2\nabla_\alpha\Big((3A^\alpha w_\nu-A^\alpha A_\nu)f_R\Big)\nonumber\\&+2\nabla_\nu(A^2f_R)+f_R\Big(4A^\alpha R_{\nu\alpha}-4w^2 A_\nu-6A^2w_\nu+2A^\alpha w_\alpha(3A_\nu+2w_\nu)-6A_\nu \nabla_\alpha A^\alpha+4(A^\alpha-w^\alpha)\nabla_\alpha A_\nu\nonumber\\&-2(w^\alpha+5A^\alpha)\nabla_\nu A^\alpha+2(A^\alpha-2w^\alpha)\nabla_\nu w_\alpha+4\Box A_\nu+2\nabla_\nu\nabla_\alpha A^\alpha+2A_\nu\nabla_\alpha w^\alpha+2(2w^\alpha-3A^\alpha)\nabla_\alpha w_\nu\Big)\nonumber\\&-4T_{\nu\alpha}\nabla^\alpha f_{T_m}+4L_m\nabla_\nu (L_mf_{T_m})-2f_{T_m}\nabla_\nu T_m\Bigg]:=f_{\nu}.
\end{align}
\end{widetext}

\subsection{The Newtonian limit}

In this subsection, we will consider the Newtonian limit of the theory. Assume that the metric can be expanded over the flat Minkowski space as
\begin{align}
ds^2=-(1+2\phi)dt^2+(1-2\psi)(dx^2+dy^2+dz^2),
\end{align}
where $\phi$ and $\psi$ are the Newtonian potentials. In the weak field limit, the only non-vanishing component of the energy-momentum tensor is $T_{00}=\rho$. For simplicity, in this Section we will assume that the Weyl and torsion vectors can be expressed as derivatives of scalar fields as
\begin{align}\label{vec}
w_\mu=\nabla_\mu w,\qquad A_\mu=\nabla_\mu A.
\end{align}
The above assumption does not restrict the theory in its Newtonian limit, since the Weyl and torsion vectors in this limit have only one dominant component, which could be rewritten as \eqref{vec} by suitable coordinate transformation.

In the weak field limit, the connection equation \eqref{affineq} is simplified to
\begin{align}\label{0}
f^0_R A-(f^0_Q+f^0_R)w+f^0_{RT_m}\rho+2f^0_{RR}(\Box\phi-2\Box\psi)=0,
\end{align}
and
\begin{align}\label{i}
f^0_Qw-2f^0_TA=0,
\end{align}
respectively. The $(00)$, $(ii)$ and $(0i)$ components of the metric equation can be written as
\begin{align}\label{00}
(f^0_R-2f^0_Q)\Box w-2f^0_R\Delta A-4f^0_R\Delta \psi+(1+f^0_{RT_m})\rho=0,
\end{align}
\begin{align}\label{ii}
4f^0_Q\Delta w+f^0_R(4\Box A-5\Box w-4\Delta\psi+4\Delta\psi)+3f^0_{T_m}\rho=0,
\end{align}
and
\begin{align}\label{0i}
f^0_R A+(f^0_Q+f^0_R)w+f^0_R(\psi-\phi)=0.
\end{align}

Now, from \eqref{i} one obtains
\begin{align}
	A=\f{f^0_Q}{2f^0_T}w,
\end{align}
which, with the help of Eq.~\eqref{0i}, gives
\begin{align}
\psi=\phi-\left(1+\f{f^0_Q}{f^0_R}+\f{f^0_Q}{2f^0_T}\right)w.
\end{align}
Using Eqs.~\eqref{00} and \eqref{ii} one obtains the generalized Poisson equation in this model as
\begin{align}
\Delta\phi=G_{eff}\rho,
\end{align}
where we have defined the generalized Newton constant (gravitational coupling) as
\begin{align}
G_{eff}=\f{1}{16\pi f^0_R}\left[1+\f13f^0_{T_m}\left(8+\left(\f{2}{f^0_R}+\f{1}{f^0_T}\right)f^0_Q\right)\right].
\end{align}
It is interesting to note that in a special case where we have no non-minimal matter couplings, i.e. $f^0_{T_m}=0$, one obtains the usual Newton constant as in GR, and we also recover the standard Poisson equation of classical gravitation.

\subsection{Thermodynamic interpretation}

An important and interesting consequence of the $f\left(R,T,Q,T_m\right)$ theory, involving
geometry-matter coupling,  is the non-conservation of the matter
energy-momentum tensor,  which is a specific property of this class of models. This property has fundamental
 physical implications, and may represent the bridge between the
interpretation of the $f\left(R,T,Q,T_m\right)$ theory as an effective classical description
of the quantum theory of gravity.

In $f\left(R,T,Q,T_m\right)$ gravity, the divergence of the matter energy-momentum tensor is given by Eq.~(\ref{cons}) in the general form  $\nabla_\alpha T^\alpha_{~\nu}=f_{\nu}$, where $f_{\nu}$ is the {\it nonconservation vector} of the theory. In the following for the matter energy-momentum tensor we adopt the perfect fluid form,
\be
T_{\mu\nu}=(\rho+p)u_{\mu}u_{\nu}+pg_{\mu \nu},
\ee
where $\rho $ and $p$ are the energy density and pressure of the fluid, while the four-velocity $u_{\mu}$ satisfies the normalization condition $u_{\mu}u^{\mu}=-1$. By multiplying Eq.~(\ref{cons} by $u^\nu$,  we obtain the  energy balance equation of the $f(R,T,Q,T_m)$ theory as
\begin{eqnarray}\label{61}
\dot{\rho}+3(\rho + p) H =u^{\nu}f_{\nu},
\end{eqnarray}
where we have denoted $H = \left( \nabla_{\mu} u^{\mu} \right)/3$, and $\dot{%
} = d/ds = u^{\mu } \nabla_{\mu }$.

We multiply now Eq.~(\ref{cons}) by the projection operator $h_{
 \lambda}^\nu$, defined as
 \be
 h_{ \lambda}^\nu \equiv
\delta_{\lambda}^\nu + u^\nu u_\lambda,
\ee
 with the properties
 \be
 u_{\nu} h_{\lambda}^{\nu} = 0, h_{\lambda}^{\nu} \nabla_{\mu} u_{\nu} = \nabla_{\mu} u_{\lambda},
\ee
and
\be
h^{\nu \lambda}
\nabla_\nu = \left( g^{\nu\lambda} + u^\nu u^\lambda \right) \nabla_\nu =
\nabla ^\lambda + u^\lambda u^{\nu} \nabla_{\nu},
\ee
respectively, we obtain the
{\it non-geodesic equation of motion} of massive test particles in $f\left(R,T,Q,T_m\right)$ theory as
\begin{eqnarray}\label{force0}
\hspace{-0.7cm}u^{\nu} \nabla_{\nu} u^{\lambda} =\frac{d^2x^{\lambda}}{d%
s^2}+\Gamma_{\mu \nu}^{\lambda }u^{\mu }u^{\nu} = \frac{- h^{\nu \lambda}
\nabla_{\nu} p + h^{\nu \lambda} f_{\nu}}{\rho + p}.
\end{eqnarray}

The non-conservation of the matter energy-momentum tensor, as well as the energy balance equation Eq.~(\ref{61}) in $f(R,T,Q,T_m)$  leads to the inescapable conclusion that, due to the presence of matter-geometry coupling, matter generation processes could
take place during the cosmological evolution. These kind of effects also
do appear in quantum field theories in curved space-times (for a detailed discussion of matter creation processes in cosmology see \cite{book}, and references therein). In quantum field theory particle creation is a direct consequence of time varying
gravitational field. Hence, $f\left(R,T,Q,T_m\right)$ theory, involving particle production, could also lead to a semiclassical description of
quantum processes taking place in gravitational fields.

\subsubsection{Thermodynamic quantities in presence of matter creation}

The fact that the divergence of the matter energy-momentum tensor is different from zero can be interpreted as indicating the presence of particles creation. In the presence of matter production all the basic equilibrium quantities, including the particle and entropy fluxes, must be modified to
include particle creation \cite{P-M,Lima,Su}. The balance equation for the particle flux
$N^{\mu} \equiv nu^{\mu}$, where $n$ is the particle number density, must be modified in the presence of particle creation as
\begin{equation}
\nabla _{\mu}N^{\mu}=\dot{n}+3Hn=n\Gamma,
\end{equation}
where $\Gamma $ is the particle creation rate. If $\Gamma \ll H$, particle creation effects can be neglected. The entropy flux vector is given by $S^{\mu} \equiv \tilde{s}u^{\mu} = n\sigma u^{\mu}$, where $\tilde{s}$ is the entropy density, while by $%
\sigma $ we have denoted the entropy per particle. For the divergence of the entropy flux we find the expression
\begin{equation}\label{62b}
\nabla _{\mu}S^{\mu}=n\dot{\sigma}+n\sigma \Gamma\geq 0.
\end{equation}
For $\sigma ={\rm constant}$, we obtain
\begin{equation}
\nabla _{\mu}S^{\mu}=n\sigma \Gamma =\tilde{s}\Gamma \geq 0.
\end{equation}

Hence, the variation of the entropy is fully determined by the particle creation processes from the gravitational field. Moreover,the condition  $\tilde{s}>0$ implies that the particle creation rate $\Gamma$ must satisfy the condition $\Gamma \geq 0$, implying that the gravitational fields could create particles, with the opposite process forbidden.  In the presence of matter production the
energy-momentum tensor must to include the second law of thermodynamics, which can be implemented by considering a correction to the equilibrium expression of the energy-momentum tensor $T^{\mu \nu}_{\text{eq}}$, so that \cite{Bar}
\begin{equation}\label{64}
T^{\mu \nu}=T^{\mu \nu}_\text{eq}+\Delta T^{\mu \nu},
\end{equation}
where $\Delta T^{\mu \nu}$ is a new term due to
particle production, satisfying the conditions
\begin{equation}
\Delta T_{\; 0}^0=0, \quad \Delta T_{\; i}^j=-P_c\delta_{\; i}^j,
\end{equation}
where $P_c$ is called the \textit{creation pressure}, and it describes
the  effects of particle production from the gravitational field in a
macroscopic system. Hence,
\begin{equation}
\Delta T^{\mu \nu}=-P_ch^{\mu \nu}=-P_c\left(g^{\mu \nu}+u^{\mu}u^{\nu}\right),
\end{equation}
so that $u_{\mu}\nabla _{\nu}\Delta T^{\mu \nu}=3HP_c$.
Thus the total energy balance
equation $u_{\mu}\nabla _{\nu}T^{\mu \nu}=0$, which follows from Eq.~\eqref{64}, can be reformulated as
\begin{equation}
\dot{\rho}+3H\left(\rho+p+P_c\right)=0.
\end{equation}

Another important thermodynamic relation, the Gibbs law, given by
formulated as \cite{Lima}
\begin{equation}
n \mathcal{T} \mathrm{d} \left(\frac{s}{n}\right)=n\mathcal{T}\mathrm{d}\sigma=\mathrm{d}\rho -\frac{\rho+p}{n}\mathrm{d}n,
\end{equation}
must also be satisfied by the thermodynamic quantities, where by $\mathcal{T}$ we have denoted the thermodynamic temperature of the system.

\subsubsection{Particle creation in $f\left(R,T,Q,T_m\right)$ theory}

By using some simple algebraic transformations the energy balance equation in $f\left(R,T,Q,T_m\right)$ theory, Eq.(~\eqref{61}), can be reformulated as
\begin{equation}\label{76}
\dot{\rho}+3H\left( \rho +p+P_{c}\right) =0,
\end{equation}%
where the creation pressure $P_{c}$ is given by
\be
P_c=-\frac{u_{\nu}f^{\nu}}{3H}.
\ee

Hence, the energy balance equation Eq.~\eqref{61} can be obtained from the covariant divergence of the effective energy momentum tensor $%
T^{\mu \nu }$, defined according to
\begin{equation}
T^{\mu \nu }=\left( \rho +p+P_{c}\right) u^{\mu }u^{\nu }+\left(
p+P_{c}\right) g^{\mu \nu }.
\end{equation}

From the Gibbs law, by assuming that particle creation is an adiabatic process, with $\dot{\sigma}=0$, we obtain
\begin{equation}
\dot{\rho}
=\left(\rho+p\right)\frac{\dot{n}}{n}
=\left(\rho+p\right)\left(\Gamma-3H\right).
\end{equation}
With the use of the energy balance equation we obtain the
relation between the matter production rate $\Gamma $ and the creation pressure $P_c$ as
\begin{equation}
\Gamma=\frac{-3HP_c}{\rho+p}.
\end{equation}

The divergence of the entropy flux vector is given by
\begin{equation}
\nabla _{\mu}S^{\mu}=\frac{-3 n \sigma H P_c}{\rho+p}.
\end{equation}

Due to their complexity, we will not present the explicit expressions of the thermodynamic parameters describing matter creation in $f\left(R,T,Q,T_m\right)$ theory. The matter creation rate, as well as the creation pressure is a complicated expression involving both the Weyl and the torsion tensors, as well as their covariant derivatives. In order to gain a deeper understanding into this problem some particular cases of the theory must be first considered in full detail.

%%%%%%%%%%%%%%%%%%%%%%%%%%%%%%%%%%%%%%
\section{Cosmological implications}\label{sect3}
%%%%%%%%%%%%%%%%%%%%%%%%%%%%%%%%%%%%%%%

In the present Section we consider some cosmological implications of the $f\left(R,T,Q,T_m\right)$ theory. For the metric of the Universe we adopt the  Friedmann-Lemaitre-Robertson-Walker (FLRW) form, which in {\it conformal coordinates} takes the form
 \begin{equation}
 ds^2=a^2(t)\left(-dt^2+d\vec{x}^2\right),
 \end{equation}
 where $a=a(t)$ is the conformal scale factor. We also introduce the Hubble parameter, defined according to $H=\dot{a}/a$. We assume that the matter sector of the Universe can be described by a perfect fluid, whose equations of motion can be obtained from the Lagrangian $L_m=-\rho$, and which has  the energy-momentum tensor,
 \begin{align}
 T^\mu_{~\nu}=\textmd{diag}(-\rho,p,p,p),
 \end{align}
where $\rho$ and $p$ are the energy density and the thermodynamic pressure, respectively.

\subsection{The generalized Friedmann and Raychaudhury equations}

Since the space-time is homogeneous and isotropic, the Weyl and torsion vectors can be written as
\begin{align}
A_{\mu}=(aA_0,\vec{0}),\nonumber\\
w_\mu=(aw_0,\vec{0}),
\end{align}
where $A_0=A_0(t)$ and $w_0=w_0(t)$ are two arbitrary functions of time. In this case, the equation of motion of the affine connection Eq.~\eqref{affineq} has two independent components, corresponding to the Weyl, and torsion vectors, which can be simplified as
\begin{align}\label{affineeqcos}
&f_R(A_0-w_0)-f_Q w_0=\frac{1}{a}\dot{f}_R,\nonumber\\
&2f_TA_0-f_Qw_0=0.
\end{align}

Then the generalized Friedman and Raychaudhuri equations can be obtained as \cite{M1a}
\bea\label{metreqcos1}
&&\frac12a^2\Big[f-12A_0^2f_T+3A_0w_0f_R\Big] \nonumber\\
&&+\frac32a\Big[\dot{w}_0+Hw_0-2H(w_0f_Q+A_0f_R)\Big]-3\dot{H}f_R=\frac12a^2\rho,\nonumber\\
\eea
and
\bea\label{metreqcos2}
&&a^2f-2(\dot{H}+2H^2)f_R+2a^2(\rho+p)f_{T_m}+2aw_0\dot{f}_Q\nonumber\\
&&-a\Big[2\dot{w}_0+w_0(4H+2aw_0-3aA_0)\Big]f_Q \nonumber\\
&&+a\Big[\dot{w}_0-2\dot{A}_0+a(2A_0-w_0)(A_0+w_0)\nonumber\\
&&-(4A_0-5w_0)H\Big]f_R=-a^2p,
\eea
respectively.

\subsection{The linear case: $f=\kappa^2(R-2\Lambda)+\beta T+\gamma Q+\sigma T_m$}

In this Subsection, we will study the cosmological implications of the linear model with Lagrange density given by \cite{M1a}
$$f=\kappa^2(R-2\Lambda)+\beta T+\gamma Q+\sigma T_m.$$

In this case the equations of motion of the affine connection, Eq.~ \eqref{affineeqcos}, simplify to
\begin{align}
&\kappa^2A_0-(\gamma+\kappa^2)w_0=0,\nonumber\\
&2\beta A_0-\gamma w_0=0,
\end{align}
which implies that both Weyl and torsion vectors vanishes $w_0=0=A_0$. As a result, the linear model does not contain torsion and non-metricity. Using the above results, one can rewrite the metric equations \eqref{metreqcos1} and \eqref{metreqcos2} as
\begin{align}
6\kappa^2 H^2&=a^2\Big[(1+\sigma)\rho-3\sigma p+2\kappa^2\Lambda\Big],\nonumber\\
2\kappa^2(2\dot{H}+H^2)&=-a^2\Big[\sigma\rho+(5\sigma+1)p-2\kappa^2\Lambda\Big].
\end{align}

Let us introduce now the set of the dimensionless variables $\left(\tau, h, \bar{\rho}, \Omega_{\Lambda}\right)$, given by
\begin{align}
\tau&=H_0t, H=H_0h,
\bar{\rho}=\frac{\rho}{6\kappa^2 H_0^2}, \Omega_{\Lambda}=\frac{\Lambda}{3H_0^2}.
\end{align}

Here $H_0$ is the present day value of the Hubble parameter. In the following we will also assume that the Universe is filled with pressureless dust, consisting of non-interacting particles with matter density $\bar{\rho}_m$. This choice of the matter content does not affect in fact the late time cosmological behavior of the model, since the energy density of the electromagnetic radiation component has very small values today. In order to compare the result with the observations it is convenient to transform the field equations by introducing the redshift independent variable. The redshift $z$ is defined according to
\begin{align}
1+z=\frac{1}{a}.
\end{align}
Then the generalized Friedmann equation of the model can be written as
\begin{align}
(1+z)^2h^2=(1+\sigma)\bar{\rho}_m+\Omega_{\Lambda}.
\end{align}

From the definition of the dimensionless Hubble parameter $h(z=0)=1$, one obtains density parameter corresponding to the cosmological constant as
\begin{align}
\Omega_{\Lambda}=1-(1+\sigma)\Omega_{m0}.
\end{align}

Also, from the Raychaudhuri equation we obtain the differential equation satisfied by the dimensionless matter density as
\begin{align}
(1+\sigma)(1+z)\bar\rho_m^\prime-3(1+2\sigma)\bar\rho_m=0,
\end{align}
with the general  solution given by
\begin{align}
\bar\rho_m=\Omega_{m0}(1+z)^{\frac{3(1+2\sigma)}{1+\sigma}},
\end{align}
where $\Omega_{m0}$ is the current value of the dust density abundance.

\subsubsection{Fitting the model with the cosmological observations}

In the following, we will estimate the best fit values of the model parameters $H_0$, $\Omega_{m0}$ and $\sigma$ by using the observational data on the Hubble parameter in the redshift range $0<z<2$, as presented in \cite{hubble}. To perform this comparison, we use the likelihood analysis of the model based on the data on $H_0$. In the case of independent data points, the {\it likelihood function} is defined according to
\begin{align}
L=L_0e^{-\chi^2/2},
\end{align}
where $L_0$ is the normalization constant, and the quantity $\chi^2$ is given by
\begin{align}
\chi^2=\sum_i\left(\frac{O_i-T_i}{\sigma_i}\right)^2.
\end{align}
Here $i$ indicates the number of the data points, $O_i$ represent the observational values, $T_i$ are the values obtained from the theoretical model, and $\sigma_i$ are the observational errors associated with the $i$th data point.
For the linear theory, the likelihood function can be defined as
\begin{align}
L=L_0\,\textmd{exp}\left[-\frac12\sum_i\left(\frac{O_i-\sigma_8^0T_i}{\sigma_i}\right)^2\right],
\end{align}

By maximizing the likelihood function one can find the best fit values of the parameters. In Table \ref{tab1}, we have summarized the result of the maximum likelihood estimation on the parameters $H_0$, $\Omega_{m0}$ and $\sigma$ together with their $1\sigma$ and $2\sigma$ confidence interval.
\begin{table}[h]
	\begin{center}
		\begin{tabular}{|c||c|c|c|}
			\hline
			Parameter&Best fit value&$1\sigma$~interval&$2\sigma$~interval\\
			\hline
			$H_0$&$66.22$&$66.22\pm3.38$&$66.22\pm6.62$\\
			\hline
			$\Omega_{m0}$&$0.45$&$0.45\pm0.202$&$0.45\pm0.39$\\
			\hline
			$\sigma$&$-0.60$&$-0.60\pm0.09$&$-0.60\pm0.17$\\
			\hline
		\end{tabular}
	\end{center}
	\caption{Best fit values of the linear model parameters $H_0$, $\Omega_{m0}$ and $\sigma$ together with their $1\sigma$ and $2\sigma$ confidence intervals.\label{tab1}}
\end{table}

\begin{figure*}[ht]
	\centering
	\includegraphics[scale=0.57]{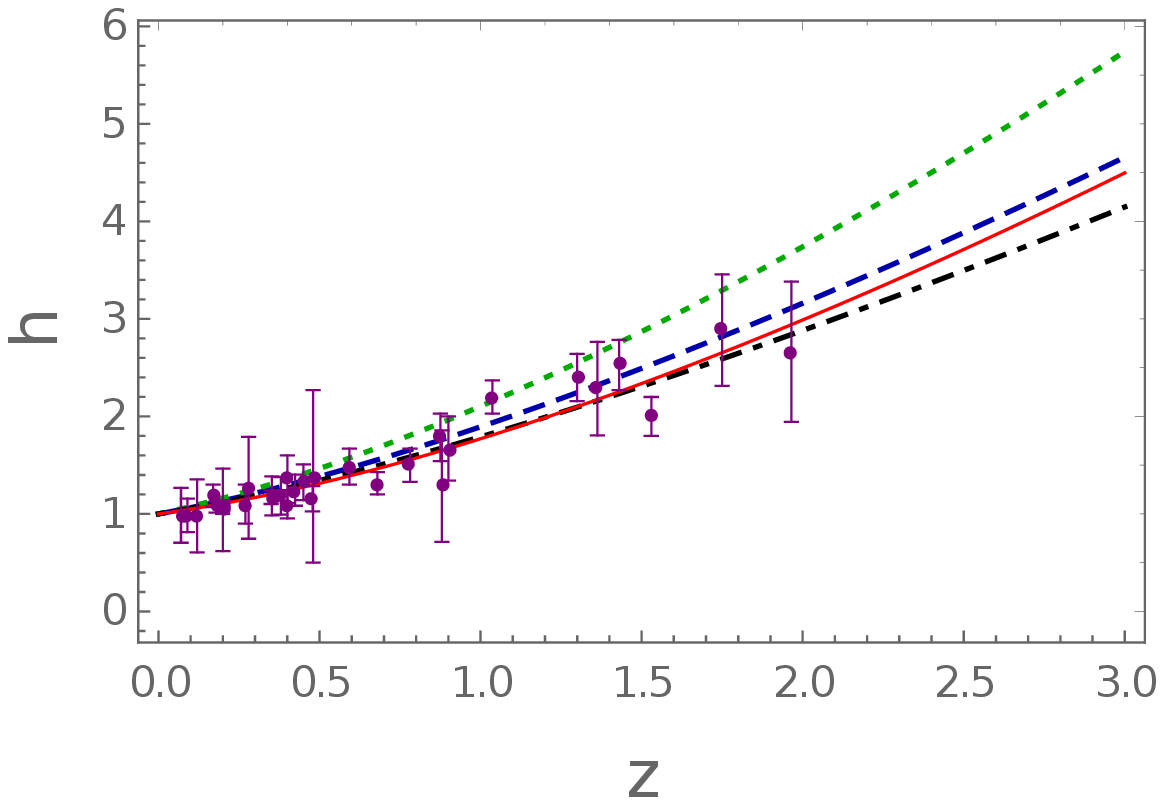}\includegraphics[scale=0.57]{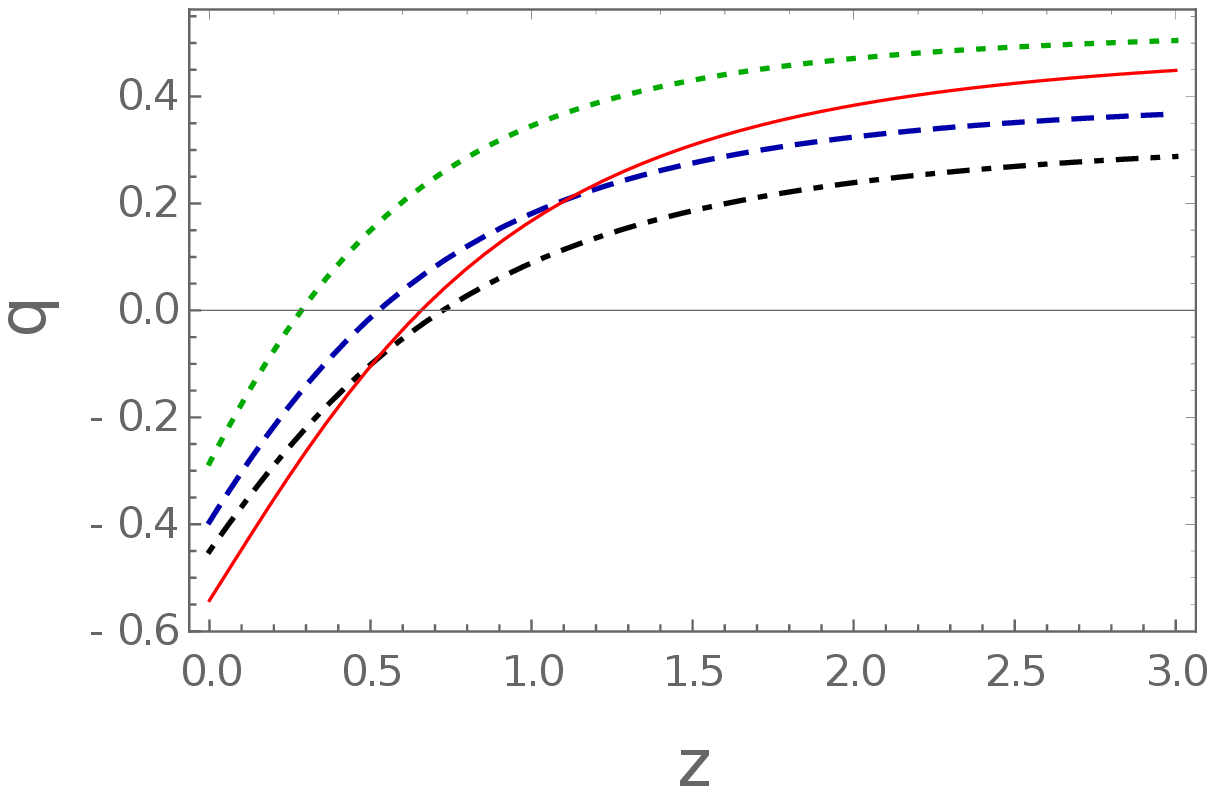}
	\caption{The evolution of the Hubble function $(1+z)h$  (left panel), and of the deceleration parameter $q$ (right panel)  as a function of redshift for different values of $\sigma$: $\sigma=-0.06$ (best fit, dashed), $0.02$ (dotted), and  $-0.1$ (dot-dashed), respectively. To obtain the plots we have used the best fit values for the Hubble parameter. The solid red line corresponds to the $\Lambda$CDM model.  The error bars indicate the observational values \cite{hubble}.}\label{fig1}
\end{figure*}

The evolution of the Hubble function, of the deceleration parameter, and of the matter density are given, as a function of the redshift, by
\begin{align}
q=(1+z)\frac{h^\prime}{h},\quad \Omega_m=\frac{\bar\rho_m}{h^2(1+z)^2},
\end{align}
and they presented in Fig.~\ref{fig1}.

As one can see from Fig.~(\ref{fig1}), the simple linear model of the $f\left(R,Q,T,T_m\right)$ theory can give an acceptable description of the cosmological observations of the Hubble function up to a redshift of $z=2$, and, for some model parameters, even up to a redshift of $z=3$. There is also a good concordance with the predictions of the $\Lambda$CDM model. On the other hand, important differences do appear in the behavior of the deceleration parameter between the present and the $\Lambda$CDM models, at both low and high redshifts. However, both models predict a very similar value for the transition from deceleration to acceleration. The differences in the present day value of $q$ may allow to test the validity of each model.

\begin{figure}[ht]
\centering
\includegraphics[scale=0.57]{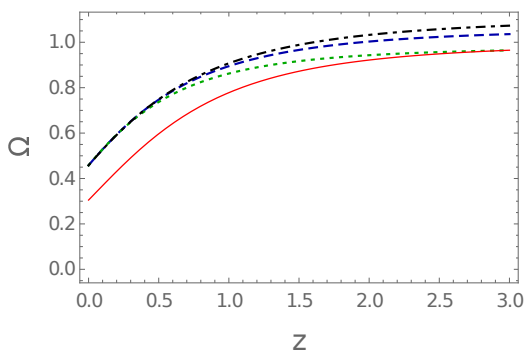}
\caption{\label{fig1a}The evolution of the matter energy density  as a function of the redshift for different values of $\sigma$: $\sigma=-0.06$ (best fit, dashed), $0.02$ (dotted), and $-0.1$ (dot-dashed), respectively. In the plots we have used the observational value of the present day matter energy density. The solid red line corresponds to the predictions of the $\Lambda$CDM model. }
\end{figure}

The evolution of the matter energy density is represented in Fig.~\ref{fig1a}.

There are important quantitative differences between the predictions of the two models, with the linear $f\left(R,Q,T,T_m\right)$ model giving higher matter density values at the present time, as compared to the $\Lambda$CDM model. Hence, exact astronomical and astrophysical determinations of the matter density evolution could provide an alternative test of the $\Lambda$CDM model, and of modified gravity theories.

\subsection{The multiplicative case: $f=\kappa^2(R-2\Lambda)-\frac13\beta^2T Q+\sigma T_m$}

In this subsection we will consider the case where the torsion and non-metricity scalars couple non-minimally in the action. Hence, we assume that the Lagrangian density of the gravitational field can be obtained as
$$f=\kappa^2(R-2\Lambda)-\frac13\beta^2 T Q+\sigma T_m.$$

The equations of motion of the affine connection can be simplified in this case as
\begin{align}
&\kappa^2(A_0-w_0)+2\beta^2A_0^2w_0=0,\nonumber\\
&2\beta^2A_0w_0(2A_0-w_0)=0,
\end{align}
with the solution
\begin{align}
A_0=\frac{\kappa}{2\beta},\qquad w_0=\frac{\kappa}{\beta}.
\end{align}

Using the above expressions for $w_0$ and $A_0$,  the generalized Friedmann and Raychaudhuri equations become
\bea
6\kappa^2H^2-\frac{3\kappa^2}{2\beta^2}a(\kappa a+\beta H)&=&a^2(1+\sigma)\rho+2\kappa^2\Lambda a^2\nonumber\\
&&-3\sigma a^2 p,
\eea
and
\bea
\hspace{-0.5cm}2\kappa^2(H^2+2\dot{H})&=&-(1+5\sigma)a^2 p-\sigma a^2\rho \nonumber\\
\hspace{-0.5cm}&&+\frac{\kappa^3}{4\beta^2}a(5\kappa a-2\beta H)+2\kappa^2\Lambda a^2,
\eea
respectively.

\subsubsection{Fitting with the observational results}

Defining a dimensionless constant $\gamma=4H_0\beta/\kappa$, and transforming to the redshift variables, one can obtain the dimensionless Hubble function as
\begin{align}
2\gamma(1+z)h=1+\sqrt{4\gamma^2\big[\Omega_\Lambda+(1+\sigma)\bar\rho_m\big]-15}.
\end{align}

From the definition of the dimensionless Hubble function we obtain the density parameter associated to the cosmological constant  as
\begin{align}
\Omega_{\Lambda}=1-(1+\sigma)\Omega_{m0}+\frac{4-\gamma}{\gamma^2}.
\end{align}

The evolution equation of the dimensionless matter density $\bar\rho_m$ can be obtained as
\bea
&&2\gamma(1+z)h\left[(1+\sigma)(1+z)\bar\rho_m^\prime-3(1+2\sigma)\bar\rho_m+\frac{7}{\gamma^2}\right]\nonumber\\
&&-(1-2\sigma)\bar\rho_m-4\Omega_{\Lambda}+\frac{8}{\gamma^2}=0.
\eea

Similarly to the previous case, we will find the best fit values of the parameters $H_0$, $\sigma$, $\gamma$ and $\Omega_{m0}$, respectively,  using the observational data on the Hubble parameter. The results are summarized in Table \eqref{tab2}.
\begin{table}[h]
	\begin{center}
		\begin{tabular}{|c||c|c|c|}
			\hline
			Parameter&Best fit value&$1\sigma$~interval&$2\sigma$~interval\\
			\hline
			$H_0$&$67.42$&$67.42\pm1.41$&$67.42\pm2.76$\\
			\hline
			$\Omega_{m0}$&$0.33$&$0.33\pm0.024$&$0.33\pm0.047$\\
			\hline
			$\sigma$&$0.0012$&$0.0012\pm0.023$&$0.0012\pm0.046$\\
			\hline
			$\gamma$&$21.88$&$21.88\pm28.73$&$21.88\pm56.31$
			\\
			\hline
		\end{tabular}
	\end{center}
	\caption{Best fit values of the multiplicative model parameters $H_0$, $\Omega_{m0}$, $\gamma$ and $\sigma$ together with their $1\sigma$ and $2\sigma$ confidence intervals.\label{tab2}}
\end{table}

\begin{figure*}
	\centering
	\includegraphics[scale=0.57]{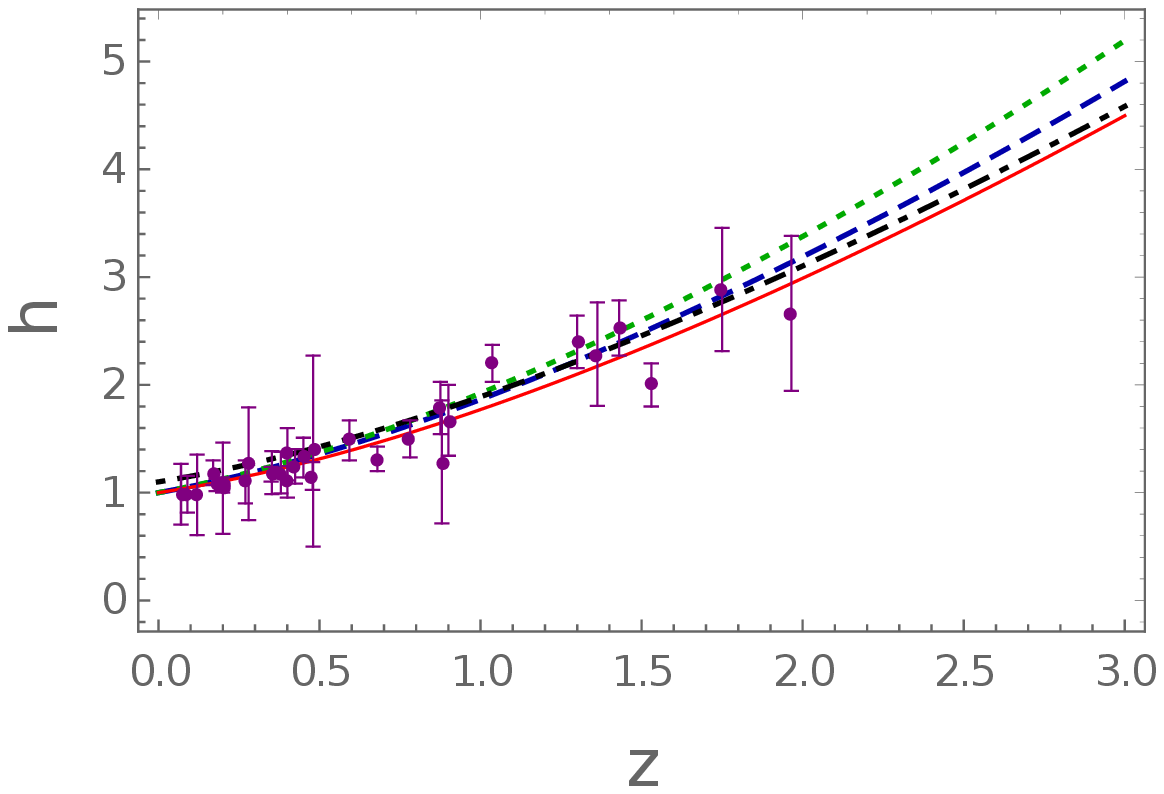}\includegraphics[scale=0.57]{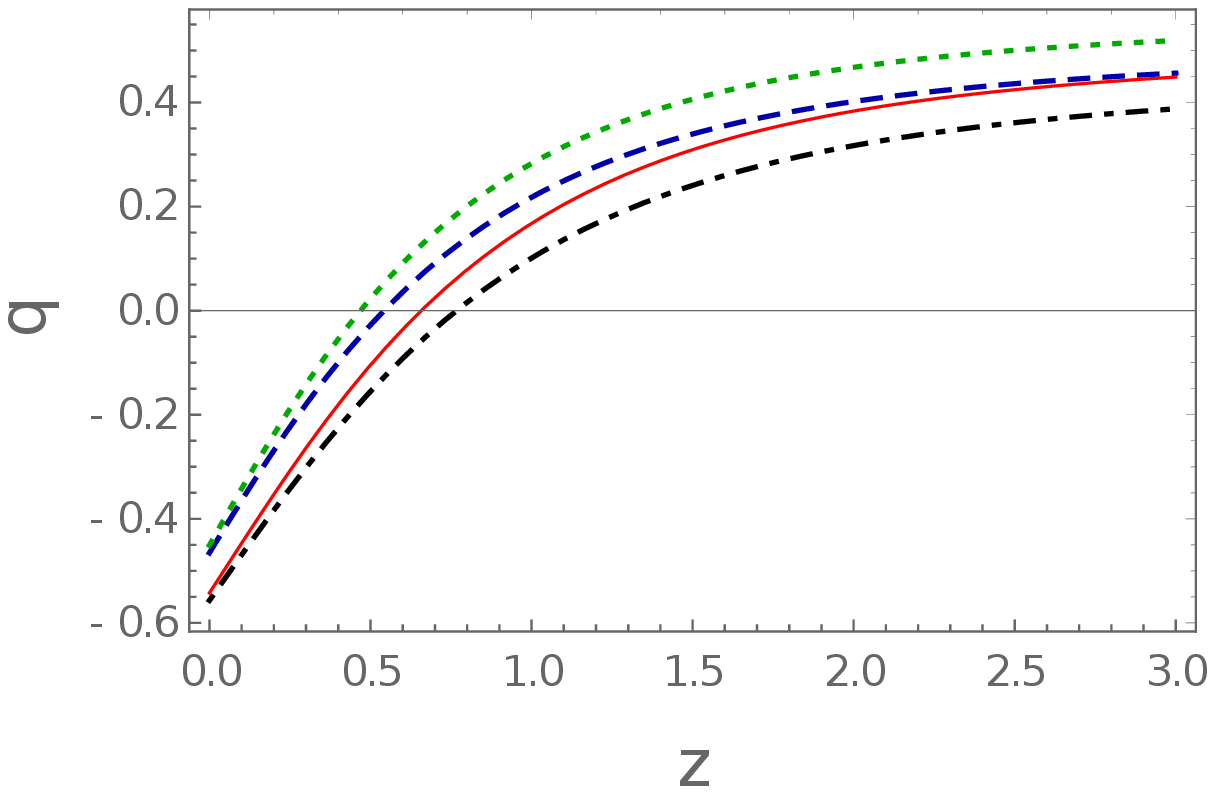}
	\caption{Evolutions of the Hubble function $(1+z)h$ (left panel) and of the deceleration parameter $q$ (right panel) as a function of the redshift for different values of  $(\sigma,\gamma)$: $(\sigma,\gamma)=(0.0012,21.88)$ (best fit, dashed), $(0.04,70)$ (dotted), and $(-0.03,-10)$ (dot-dashed), respectively. We have used the best fit values for the Hubble parameter and the current matter density abundance. The solid red line corresponds to the $\Lambda$CDM theory.  The error bars indicate the observational values \cite{hubble}.}\label{fig2}
\end{figure*}

In Fig.~ \ref{fig2}, we have plotted the evolution of the Hubble function $(1+z)h$ and of the deceleration parameter  as a function of redshift $z$.

The model gives a good description of the observational data for the Hubble function.  There are some differences in the behavior of the deceleration parameter, as compared to the $\Lambda$CDM model. Even that the present day value of $q$ can be reobtained for a specific value of the model parameters, the transition from deceleration to acceleration happen at different values of $z$ in the two models.

\begin{figure}
	\centering
	\includegraphics[scale=0.57]{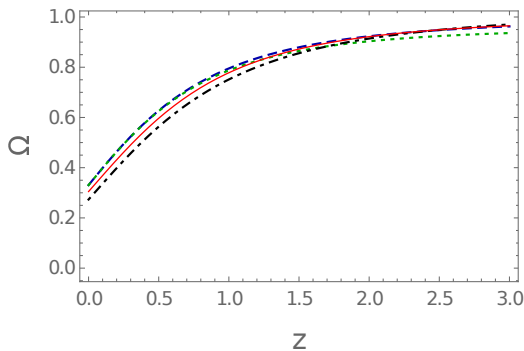}
	\caption{The evolution of the matter energy density as a function of redshift for different values of $(\sigma,\gamma)=(0.0012,21.88)$ (best fit, dashed), $(0.04,70)$ (dotted), and $(-0.03,-10)$ (dot-dashed), respectively.  The solid red line corresponds to the $\Lambda$CDM model.}\label{fig2a}
\end{figure}

The evolution of the density parameter of the ordinary matter is presented in Fig.~\ref{fig2a}.

There is a much better concordance between the predictions of the present model, and $\Lambda$CDM. However, it is worth mentioning that the multiplicative model, compared to the linear model, predicts smaller values of the present day dust energy density. However, both models predicts a current Hubble parameter of around $67$, which indicates that the present theory can not improve the Hubble tension.
\\
%%%%%%%%%%%%%%%%%%%%%%%%%%%%%%%%%%%%%%%%%%%%%%%%%%%%%%
\section{Discussions and final remarks}\label{sect4}

In the present paper we have investigated a maximal extension of the Hilbert-Einstein Lagrangian in the presence of matter, by considering a physical model, defined on a metric-affine manifold, in which the action can be expressed in terms of three geometric quantities $\left(R,T,Q\right)$-the curvature scalar, the torsion, and the nonmetricity, respectively, and one matter determined function-the trace of the matter energy-momentum tensor $T_m$, respectively. From a geometric point of view, the theory is defined in a Weyl-Cartan space-time. I our approach we have taken explicitly into account the roles played by the Weyl vector and of the torsion in the description of the gravitational phenomena. As for the matter coupling term, we have not restricted it in any way, thus allowing the possibility of couplings between curvature, torsion, and nonmetricity, respectively. The developed theory also generalizes a number of previously investigated theories, like, for example, the $f(R,T_m)$ theory  \cite{fT1}, or the $f(Q,T_m)$ theory \cite{fQC1, fQC2}.

A complete classification of all algebraic gravitational extensions of the Hilbert-Einstein action, with the matter described by the trace of the energy-momentum tensor was obtained in \cite{M1}, and it is presented in Table~\ref{class}.

\begin{center}
\begin{table}[htbp]
\begin{tabular}{||c||c||}
\hline
MGI & MGII \\
\hline
$f(R,T_m)+2\kappa ^{2}L_{m}$ & $%
f(R,Q)+2\kappa ^{2}L_{m}$ \\
\hline
MGIII & MGIV \\
\hline
$f(T_m,Q)+2\kappa ^{2}L_{m}$ & $%
f(R,T,T_m)+2\kappa
^{2}L_{m}$ \\
\hline
MGV & MGVI \\
\hline
$f(R,T,Q)+2\kappa ^{2}L_{m}$ &
$f(R,Q,T_m)+2\kappa
^{2}L_{m}$ \\
\hline
MGVII & MGVIII \\
\hline
$f(T,Q,T_m)+2\kappa
^{2}L_{m}$ & $f(R,T,Q,T_m)+2\kappa ^{2}L_{m}$ \\
\hline
\end{tabular}
\caption{Metric-affine geometric extensions of the Hilbert-Einstein action in Weyl-Cartan geometries \cite{M1}. In the Table $T_m$ denotes the trace of the energy-momentum tensor, while $T$ denotes the torsion scalar. }\label{class}
\end{table}
\end{center}

The theoretical models of Table~\ref{class} increase in complexity, with MGVIII involving all the three possible geometrical parameters, as well as {\it one single matter parameter}, the trace of the energy-momentum tensor. However, couplings generated by general matter terms involving the matter Lagrangian $L_m$, are also possible, and a first step in this direction is represented by the $f\left(R,T_m,L_m\right)$ gravity theory \cite{ZH}, which introduces a new matter degree of freedom. Some of these models have been extensively investigated in the literature \cite{M2,M3,M5}.

In the present study we have considered theoretical models in which {\it  both the nonmetricity and the torsion can be represented in terms of two vectors $\left(w_{\mu},A_{\mu}\right)$, the Weyl vector, and the torsion vector}, respectively. These choices simplify significantly the field equations, as well as the physical model, without any loss in generality. Moreover, they open the possibility of establishing some interesting connection between the two fundamental geometric quantities, nonmetricity and torsion, which allow to extend the standard Riemann geometry of general relativity. We have investigated the Newtonian limit of the theory by considering the integrable cases of the metric-affine geometries, in which both nonmetricity and torsion vectors can be represented as the covariant derivatives of a scalar potential. The weak field and low velocity limit of the theory leads to an effective Poisson equation, in which the gravitational coupling constant becomes a function of the background values of the derivatives of the gravitational Lagrangian with respect to the model variables. The Poisson equation may lead to the possibilities of testing the $f\left(R,T,Q,T_m\right)$ theory at the Solar System or even galactic scales.

We have also briefly pointed out, without fully developing it, a possible thermodynamic interpretation of the modified gravity theories with geometry -matter coupling, developed in the framework of the thermodynamic of open systems. The nonconservation of the energy-momentum tensor, as well as the energy balance equation that can be derived from it suggests that a matter/energy transfer may occur irreversibly from the gravitational field (geometry) to matter, and that this kind of theories may describe matter creation processes.

\te{Particle production is also a specific characteristic of quantum field theory in curved spacetimes. The classic example for this effect is the behavior of a real minimally coupled  scalar field $\phi$ in a cosmological geometry, with gravitational action $S=(1/2)\int{\left(g^{\mu \nu}\nabla _{\mu}\phi \nabla _{\nu}\phi-m^2\phi^2\right)\sqrt{-g}d^4x}$. By introducing the conformal time $\eta$, the metric becomes conformally Minkowskian, given by $ds^2=a^2(\eta)g_{\mu \nu}dx^{\mu}dx^{\nu}$, and the equation of motion of the field takes the form \cite{Toms}
\be
\chi '' -\nabla ^2 \chi +\left(m^2-\frac{a''}{a}\right)\chi=0,
\ee 
where $\chi =a(\eta)\phi (x,\eta)$, and the prime denotes the derivatives with respect to $\eta$. The new field $\chi $ satisfies the equation of
motion of a massive scalar field in the ordinary Minkowski spacetime, but with an effective mass $m_{eff}^2(\eta)=m^2a^2-a''/a$, which is time dependent. The time dependency of the mass leads to the interaction between the gravitational and scalar fields. Since the action for the field $\chi$ is time dependent,  the total energy of the field is not conserved. Hence, in quantum field theory
the quantization of the scalar field results in matter generation due to the presence of the classical gravitational background.}
\te{
An alternative approach to particle creation can be obtained in the framework of semiclassical gravity,  where one assumes that the matter fields are quantized, and they exist in a classical spacetime characterized by a metric $g_{\mu \nu}$, and with the gravitational fields described classically, by using the Hilbert-Einstein action, $S=\int{\left(-R/2\kappa ^2\right)\sqrt{-g}d^4x}$. In the semiclassical approach no preferred vacuum state for the fields does exist, and therefore matter generation effects occur naturally. Hence, in the presence of the gravitational fields quantum matter satisfies the semiclassical Einstein equations (see \cite{Phys} and references therein),
\be\label{1q}
R_{\mu \nu}-\frac{1}{2}g_{\mu \nu}R=\frac{8\pi G}{c^4}\left\langle\Psi\left|\hat{T}_{\mu\nu}\right|\Psi \right\rangle.
\ee
 The semiclassical Einstein  equations are obtained phenomenologically by substituting in the gravitational field equations the classical matter energy-momentum tensor $T_{\mu\nu}$, by its expectation value $\left<\Psi \right |\hat{T}_{\mu \nu}\left |\Psi \right>=T_{\mu \nu}$, where $\Psi$ describes an arbitrary quantum state. In the classical limit the matter energy-momentum tensor $T_{\mu \nu}$ is given by $\left<\Psi \right |\hat{T}_{\mu \nu}\left |\Psi \right>=T_{\mu \nu}$. }
 
\te{ The semiclassical Einstein equations ~(\ref{1q}) can also be obtained from the variational principle
$\delta \left(S_g+S_{\psi}\right)=0$ \cite{sg3a},
where $S_g$ is the general relativistic classical action of the gravitational field, while $S_{\psi}$, describing the quantum effects, is given by
\begin{equation}\label{Ki}
S_{\Psi}=\int{\left[{\rm Im}\left \langle \dot{\Psi}|\Psi\right \rangle-\left \langle \Psi |\hat{H}|\Psi \right \rangle +\alpha \left(\left \langle \Psi |\Psi \right \rangle -1\right) \right]dt},
\end{equation}
where $\hat{H}$ is the Hamilton operator of the matter, and $\alpha $ is a Lagrange multiplier.}

  The action (\ref{Ki}) can be generalized by introducing a nonminimal coupling between the quantized matter fields  and the classical Ricci scalar $R$ \cite{sg3a}. More exactly, in the simplest approach one can assume that the quantum matter-geometry coupling has the simple form
$\int{RF\left(\left<f(\phi)\right>\right)_{\Psi}\sqrt{-g}d^4x}$, where by $F$ and $f$ we have denoted two arbitrary functions. By   $\left(\left<f(\phi)\right>\right)_{\Psi}=\left<\Psi (t)\right|f[\phi (x)]\left|\Psi (t)\right>$ we have denoted the average value of an arbitrray function over the quantum fields $\Psi (t)$. With this form of the coupling, the effective semiclassical Einstein equations can be obtained as \cite{sg3a}
%\begin{eqnarray}\label{2aa}
%R_{\mu \nu}-\frac{1}{2}Rg_{\mu \nu}&=&16\pi G\Bigg[\left< \hat{T}_{\mu \nu}\right> _{\Psi}+G_{\mu \nu}F\nonumber\\
%&&-
% \nabla _{\mu}\nabla _{\nu} F+
%g_{\mu \nu}\Box F\Bigg].
%\end{eqnarray}
\be\label{2aa}
\left(1-16\pi GF\right)G_{\mu\nu}+16\pi G\left(\nabla _{\mu }\nabla _{\nu}-\Box\right)F=16\pi G \left< \hat{T}_{\mu \nu}\right> _{\Psi},
\ee
where $G_{\mu \nu}$ is the Einstein tensor, and $G$ is the gravitational constant.  

Even obtained via a quantum approach, Eq.~(\ref{2aa}) has intriguing similarities with gravitational field equations obtained in modified gravity theories. It also has an important physical consequence, which follows from the fact that the covariant divergence of the  mean value of the matter energy-momentum tensor $\left< \hat{T}^{\mu \nu}\right> _{\Psi}$ does not vanish identically, and generally $\nabla _{\mu}\left< \hat{T}^{\mu \nu}\right> _{\Psi}\neq 0$. Hence, the semiclassical quantum theoretical model described by Eq.~(\ref{2aa}) can be interpreted physically as describing particle production creation from the gravitational field. Eq.~(\ref{2aa}) may also indicate the existence of a deep (and yet unexplored) relation between modified and quantum  gravity, at least at the semiclassical level. Of course, other forms of the quantum matter-geometry couplings are possible, leading to other classes of modified gravity theories with quantum origins.

In the present paper we have investigated the theoretical basis of the particle creation processes from geometry from a classical perspective, and we have presented the general expressions for the particle creation rates, the creation pressure, and for entropy production. However, we have not presented the explicit expressions of these quantities in $f\left(R,T,Q,T_m\right)$. A detailed investigation of these effects will be performed in a new study. Moreover, it would be interesting to investigate the possibility of obtaining the $f\left(R,T,Q,T_m\right)$ action in the semiclassical approach to quantum gravity along the same lines that led to the modified gravity field equation (\ref{2aa}).  

The cosmological application of the theory have been investigated for two particular choices of the gravitational Lagrangian, corresponding to an additive and multiplicative algebraic structure of $f\left(R,T,Q,T_m\right)$. In both cases the generalized Friedmann equations have been derived, their solutions obtained numerically, and the results were compared with the cosmological observations. A comparison with the $\Lambda$CDM model was also performed. The model parameters were determined by {\it fitting the observational data for the Hubble function}. At least for a redshift of $z=1$ the model predictions are basically indistinguishable from the $\Lambda$CDM model, and fit very well the observational data. Some differences do occur at higher redshifts $z\approx 2-3$,  and these differences are more significant in the case of the simple linear additive model. These, {\it still preliminary results}, may suggest that additive/multiplicative algebraic structures of the gravitational Lagrangian, involving the products of the parameters $\left(R,T,Q,T_m\right)$ may provide a better description of cosmological phenomena than the simple additive approach.

In the present work we have considered one of the most general metric-affine gravity theory possibly to be constructed in a Weyl-Cartan geometry, which also includes geometry-matter coupling. The field equations have been derived, and some of their basic properties investigated. The cosmological results indicate that this type of theories may have the potential of explaining the present day observational data, and for opening a new window for a deeper understanding of the Universe.

%%%%%%%%%%%%%%%%%%%%%%%%%%%%%%%%%%%%%%%%%%%%
\section*{Acknowledgments}
We would like to thank to the anonymous reviewer for the careful reading of the manuscript, and for comments and suggestions that helped to improve our work. T.H. was partially supported by a grant of the Romanian Ministry of Education and Research, CNCS-UEFISCDI, project number PN-III-P4-ID-PCE-2020-2255 (PNCDI III). The work of N. M. and R. M. was  supported by the Ministry of Education and Science of the Republic of Kazakhstan, Grant  AP09058240.


\begin{thebibliography}{99}

\bibitem{Eina} A. Einstein, Sitzungsberichte der Preussischen Akademie der Wissenschaften zu Berlin, 844 (1915).

\bibitem{Hil} D. Hilbert,  Nachrichten von der Gesellschaft der Wissenschaften zu G\"{o}ttingen - Mathematisch - Physikalische Klasse {\bf 3}, 395 (1915).

\bibitem{Riem} B. Riemann,  Habilitationsschrift, 1854, Abhandlungen der K\"{o}niglichen Gesellschaft der Wissenschaften zu
G\"{o}ttingen {\bf 13}, 1 (1867).

\bibitem{Ric} G. Ricci and T. Levi-Civita,  Mathematische Annalen {\bf 54},  125 (1900).

\bibitem{Weyl}  H. Weyl, Sitzungsberichte der K\"{o}niglich Preussischen
Akademie der Wissenschaften zu Berlin {\bf 1918}, 465 (1918).

\bibitem{Car1} \'{E}. Cartan, C. R. Acad. Sci. (Paris) {\bf 174}, 593 (1922).

\bibitem{Car2}  \'{E}. Cartan, Annales \'{E}cole  Normale {\bf  40}, 325 (1923).

\bibitem{Car3} \'{E}. Cartan, Annales  \'{E}cole Normale {\bf  41}, 1 (1924).

\bibitem{Car4} \'{E}. Cartan, Annales \'{E}cole Normale {\bf 42}, 17 (1925).

\bibitem{Weit} R. Weitzenb\"{o}ck, Invariantentheorie, Noordhoff, Groningen, 1923

\bibitem{Hehl} 0] F. W. Hehl, P. von der Heyde, G. D. Kerlick, and J. M.
Nester, Review of Modern Physics {\bf 48}, 393 (1976).

\bibitem{TE1} A. Einstein, Preussische Akademie der Wissenschaften,
Phys.-math. Klasse, Sitzungsberichte {\bf 1928}, 217 (1928).

\bibitem{TE2} C. M\"{o}ller, Mat. Fys. Skr. Dan. Vid. Selsk. {\bf 1}, 10 (1961).

\bibitem{TE3} C. Pellegrini and J. Plebanski, Mat. Fys. Skr. Dan. Vid.
Selsk. {\bf 2}, 4 (1963).

\bibitem{TE4} K. Hayashi and T. Shirafuji, Phys. Rev. {\bf D 19}, 3524
(1979).

\bibitem{TE4a}  R. Ferraro and F. Fiorini,  Phys. Rev. {\bf D 75}, 084031 (2007).

\bibitem{TE4b}  S. Capozziello, O. Luongo, and E. N. Saridakis,  Phys. Rev. {\bf D 91}, 124037 (2015).

\bibitem{TE4c}  S. Capozziello, P. A. Gonzalez, E. N. Saridakis, and Y. Vasquez,  JHEP {\bf 02}, 039 (2013).
\bibitem{TE5} R. Aldrovandi and J. G. Pereira, Teleparallel Gravity,
Fundamental Theories of Physics 173, Springer, Heidelberg, 2013

\bibitem{Str} L. O'raifeartaigh and N. Straumann, Reviews of Modern Physics {\bf 72},  1 (2000).

\bibitem{Scholz} E. Scholz, arXiv:1703.03187v1 [math.HO] (2017).

\bibitem{Dirac1} P. A. M. Dirac, Proceedings Royal Society London {\bf A 333}, 403 (1973).

\bibitem{Dirac2} P. A. M. Dirac, Proceedings Royal Society London {\bf A 338}, 439 (1974).

\bibitem{Rosen} N. Rosen, Foundations of Physics {\bf 12}, 213 (1982).

\bibitem{Isrcosm} M. Israelit, Gen. Relativ. Gravit. {\bf 43}, 751 (2011).

\bibitem{Gh1} D. M. Ghilencea, JHEP {\bf 03},  049 (2019).

\bibitem{Gh2} D. M. Ghilencea and H. M. Lee, Phys. Rev. {\bf D 99}, 115007 (2019).

\bibitem{Gh3} D. M. Ghilencea, JHEP {\bf 10}, 209 (2019).

\bibitem{Gh4} D. M. Ghilencea, Phys. Rev. {\bf D 101}, 045010 (2020).

\bibitem{Gh5} D. M. Ghilencea, Eur. Phys. J. {\bf C 80}, 1147 (2020).

\bibitem{Gh6} D. M. Ghilencea, Eur. Phys. J. {\bf C 81}, 510 (2021).

\bibitem{Gh7} D. M. Ghilencea, arXiv:2104.15118 (2021).

\bibitem{Q1} J. M. Nester and H.-J. Yo, Chinese Journal of Physics
{\bf 37}, 113 (1999).

\bibitem{Q2}  J. Beltr\'{a}n Jim\'{e}nez, L. Heisenberg, and T. Koivisto,
Phys. Rev. {\bf D 98}, 044048 (2018).

\bibitem{Q3}  M. Adak, O. Sert, M. Kalay, and M. Sari, Int. J.  Mod. Phys. {\bf A 28}, (2013).

\bibitem{Q4} J. Beltr\'{a}n Jim\'{e}nez and T. S. Koivisto, Phys. Lett. {\bf B
756}, 400 (2016).

\bibitem{Q5}  A. Golovnev, T. Koivisto and M. Sandstad, Class.
Quant. Grav. {\bf 34}, 145013 (2017).

\bibitem{Q6} M. Adak, International Journal of Geometric Methods
in Modern Physics {\bf 15}, 1850198-269 (2018).

\bibitem{Q7} A. Conroy and T. Koivisto, Eur. Phys. J. {\bf C 78}, 923
(2018).

\bibitem{Q8} A. Delhom-Latorre, G. J. Olmo, and M. Ronco, Phys.
Lett. {\bf B 780}, 294 (2018).

\bibitem{Q9} L. J\"{a}rv, M. R\"{u}nkla, M. Saal, and O. Vilson, Phys. Rev.
{\bf D 97}, 124025 (2018).

\bibitem{Q10} I. Soudi, G. Farrugia, V. Gakis, J. Levi Said, and E. N.
Saridakis, Phys. Rev. {\bf D 100}, 044008 (2019).

\bibitem{Q11} M. Hohmann, C. Pfeifer, J. Levi Said, and U. Ualikhanova, Phys. Rev. {\bf D 99}, 024009 (2019).

\bibitem{Q12} K. F. Dialektopoulos, T. S. Koivisto, and S. Capozziello,
Eur. Phys. J. {\bf C 79}, 606 (2019).

\bibitem{Q13} J. Lu, X. Zhao, and G. Chee, Eur. Phys. J. {\bf C 79}, 530
(2019).

\bibitem{Q14} R. Lazkoz, F. S. N. Lobo, M. Ortiz-Ba${\rm \tilde{n}}$o and V.
Salzano, Phys. Rev. {\bf D 100}, 104027 (2019).

\bibitem{Q15} J. Beltr\'{a}n Jim\'{e}nez, L. Heisenberg, and T. S. Koivisto,
Universe {\bf 5}, 173 (2019).

\bibitem{Q16} J. Beltr\'{a}n Jim\'{e}nez, L. Heisenberg, T. S. Koivisto, and
S. Pekar, Phys. Rev. {\bf D 101}, 103507 (2020).

\bibitem{Q17} J. Beltr\'{a}n Jim\'{e}nez, L. Heisenberg, D. Iosifidis, A.
Jim\'{e}nez-Cano, and T. S. Koivisto, Phys. Lett. {\bf B 805},
135422 (2020).

\bibitem{Q18} W. Khyllep, A. Paliathanasis, and J. Dutta, Phys. Rev. {\bf D 103}, 103521 (2021).

\bibitem{Q19} R.-H. Lin and X.-H. Zhai, Phys. Rev. {\bf D 103}, 124001 (2021).

\bibitem{WC1}  H.-H. von Borzeszkowski and H.-J. Treder, Gen. Rel. Grav. {\bf 29},  455 (1997).
\bibitem{WC2} D. Puetzfeld and R. Tresguerres, Class. Quant.
Grav. {\bf 18},  677 (2001).

\bibitem{WC3} D. Putzfeld, Class. Quant. Grav. {\bf 19},  4463 (2002).

\bibitem{WC4}  D. Puetzfeld, Clas. Quant. Grav. {\bf 19}, 3263 (2002).

\bibitem{WC5} O. V. Babourova, Grav. Cosm. {\bf 10},  121 (2004).

\bibitem{WC6}  O. V. Babourova and V. F. Korolev, Russian Physics Journal {\bf 49}, 628 (2006).

\bibitem{WC7}  O. V. Baburova, V. Ch. Zhukovsky, and B. N. Frolov, Theoretical and Mathematical Physics {\bf 157},  1420 (2008).


\bibitem{WC8} T. Y. Moon, J. Lee, and P. Oh, Mod. Phys. Lett. {\bf A 25}, 3129 (2010).

\bibitem{WC9}  T. Y. Moon, P. Oh, J. S. Sohn, JCAP {\bf 11}, 005 (2010).

\bibitem{Rev} M. Novello and S. E. Perez Bergliaffa, Physics Reports {\bf 463} 127  (2008).

\bibitem{1n} A. G. Riess et al., Astron. J. \textbf{116}, 1009 (1998).

\bibitem{2n} S. Perlmutter et al., Astrophys. J. \textbf{517}, 565 (1999).

\bibitem{3n} R. A. Knop et al., Astrophys. J. \textbf{598}, 102 (2003).

\bibitem{4n} R. Amanullah et al., Astrophys. J. \textbf{716}, 712 (2010).

\bibitem{acc} D. H. Weinberg, M. J. Mortonson, D. J. Eisenstein, C. Hirata,
A. G. Riess, and E. Rozo, Physics Reports \textbf{530}, 87 (2013).

\bibitem{Ein} A. Einstein, Sitzungsberichte der K\"{o}niglich Preussischen Akademie der Wissenschaften, Berlin,  part 1: 142 (1917).

\bibitem{Sal} P. Salucci, N. Turini, and C. Di Paolo, Universe {\bf  6},  118 (2020).

\bibitem{C1}  S. Alam et al. (BOSS Collaboration),  Mon. Not. R. Astron. Soc. {\bf 470}, 2617 (2017).
1607.03155

\bibitem{C2}  T. M. C. Abbott et al. (DES Collaboration), Phys. Rev. {\bf D 98}, 043526 (2018).

\bibitem{C3} M. Tanabashi et al. (Particle Data Group),
Phys. Rev. {\bf D 98}, 030001  2018.

\bibitem{C4}  N. Aghanim et al. (Planck Collaboration), Astron. Astrophys. {\bf 641}, A6 (2020).

\bibitem{Wein1} S. Weinberg, Reviews of Modern Physics {\bf 61}, 1 (1989).

\bibitem{Wein2} H. Martel, P. R. Shapiro, and S. Weinberg, Astrophys. J. {\bf 492},  29 (1998).

\bibitem{Wein3} S. Weinberg, The Cosmological Constant Problems, in Sources and Detection of Dark Matter and Dark Energy in the Universe. Fourth International Symposium, held February 23-25, 2000, at Marina del Rey, California, USA, David B. Cline, Editor, Springer-Verlag, Berlin, New York, p.18, 2001;  arXiv:astro-ph/0005265v1

\bibitem{Lake} M. J. Lake, 	arXiv:2005.12724 [gr-qc] (2020).

\bibitem{M1} A. G. Riess, S. Casertano, W. Yuan, L. M. Macri,
and D. Scolnic,  Astrophys. J. {\bf 876}, 85 (2019).

\bibitem{M2} C. D. Huang, A. G. Riess, W. Yuan, L. M. Macri, N. L.
Zakamska, S. Casertano, P. A. Whitelock, S. L. Hoffmann, A. V. Filippenko, and D. Scolnic,
The Astrophysical Journal {\bf  889}, 5 (2020).

\bibitem{M3}  D. Pesce et al.,  Astrophys. J. Lett. {\bf 891}, L1 (2020).

\bibitem{HL20} T. Harko and F. S. N. Lobo, Int. J. Mod. Phys. {\bf D 29}, 2030008 (2020).

\bibitem{Qa1} C. Wetterich, Nuclear Physics {\bf B 302}, 645 (1988).
\bibitem{Qa2} P. J. E. Peebles and B. Ratra, Astrophys. J. Lett. {\bf 325},
L17 (1988).
\bibitem{Qa3}  B. Ratra and P. J. E. Peebles, Phys. Rev {\bf D 37}, 3406
(1988).
\bibitem{Qa4} R. R. Caldwell, R. Dave, and P. J. Steinhardt, Phys.
Rev. Lett. {\bf 80}, 1582 (1998).
\bibitem{Qa5}  L. Amendola, Phys. Rev. {\bf D 62}, 043511 (2000).

\bibitem{Qa6} A. Banerjee, H. Cai, L. Heisenberg, E. \'{O} Colg\'{a}in, M. M. Sheikh-Jabbari, and T. Yang, Phys. Rev. {\bf D 103}, L081305 ( 2021).

\bibitem{K1} C. Armendariz-Picon, V. F. Mukhanov, and P. J. Steinhardt, Phys. Rev. Lett. {\bf 85}, 4438 (2000).

\bibitem{K2} T. Chiba, T. Okabe, and  M. Yamaguchi, Phys. Rev. {\bf D 62}, 023511 (2000).

\bibitem{K3} C. Armendariz-Picon, V. F. Mukhanov and P. J. Steinhardt, Phys. Rev. {\bf D 63}, 103510 (2001).

\bibitem{T1}  A. Sen, JHEP 0207, 065 (2002).

\bibitem{T2} A. Mohammadi, T. Golanbari, H. Sheikhahmadi, K. Sayar, L. Akhtari, M. A. Rasheed and K. Saaidi, Chinese
Physics {\bf C 44}, 095101 (2020).

\bibitem{Ph1} R. R. Caldwell, Phys. Lett. {\bf B 545}, 23 (2002).

\bibitem{Ph2}  R. R. Caldwell, M. Kamionkowski and N. N. Weinberg,
Phys. Rev. Lett. {\bf 91}, 071301 (2003).

\bibitem{Ph3}  J. M. Cline, S. Y. Jeon and G. D. Moore, Phys. Rev. {\bf D 70}, 043543 (2004).

\bibitem{Qu1} E. Elizalde, S. Nojiri and S. D. Odinstov, Phys. Rev. {\bf D 70}, 043539 (2004).

\bibitem{Qu2}  S. Nojiri, S. D. Odintsov and S. Tsujikawa, Phys. Rev.
{\bf D 71}, 063004 (2005).

\bibitem{Qu3} A. Anisimov, E. Babichev and A. Vikman, J. Cosmol.
Astropart. Phys. {\bf 06}, 006 (2005)

\bibitem{Ch1} J. Khoury and A. Weltman, Phys. Rev. {\bf D 69}, 044026 (2004).
\bibitem{Ch2}  Kh. Saaidi, A. Mohammadi, and H. Sheikhahmadi, Phys. Rev. {\bf D 83}, 104019 (2011).
\bibitem{Ch3} Kh. Saaidi, and A. Mohammadi, Phys. Rev. {\bf D 85}, 023526 (2012).
\bibitem{nonlocal} Z. Haghani, arXiv:1911.06990 [gr-qc].
\bibitem{Ch4}  Kh. Saaidi, A. Mohammadi, T. Golanbari, H. Sheikhahmadi and B. Ratra, Phys. Rev. {\bf D 86}, 045007 (2012).
\bibitem{Cha1} A. Yu. Kamenshchik, U. Moschellai, V. Pasquier, Phys.
Lett. {\bf B 511}, 265 (2001).
\bibitem{Cha2} M. C. Bento, O. Bertolami, A. A. Sen, Phys. Rev. {\bf D 66}, 043507 (2002).

\bibitem{V1} C. G. Boehmer and T. Harko, Eur. Phys. J. {\bf C 50}, 423 (2007).

\bibitem{V2} Z. Haghani, T. Harko, H. R. Sepangi, and S. Shahidi, Eur. Phys. J. {\bf C  77}, 137 (2017).

\bibitem{V3} Z. Haghani, T. Harko, and S. Shahidi, Physics of the Dark Universe {\bf  21},  27 (2018).

\bibitem{Rev1} A. Joyce, B. Jain, J. Khoury, and M. Trodden,  Phys. Rept. {\bf 568},  1 (2015).

\bibitem{Rev2} A. Joyce, L. Lombriser, and F. Schmidt, Annu. Rev. Nucl. Part. Sci. {\bf 66}, 95 (2016).

\bibitem{Rev3} A. N. Tawfik and E. A. El Dahab, Gravitation and Cosmology {\bf  25}, 103 (2019).

\bibitem{Rev4} N. Frusciante and L. Perenon, Phys. Rept. {\bf 857},  1 (2020).

\bibitem{Bu1} H. A. Buchdahl, Mon. Not. Roy. Astron. Soc. \textbf{150}, 1
(1970).

\bibitem{Bu2} R. Kerner, Gen. Rel. Grav. \textbf{14}, 453 (1982).

\bibitem{Bu3} J. P. Duruisseau, R. Kerner and P. Eysseric, Gen. Rel. Grav. \textbf{15}, 797
(1983).

\bibitem{Bu4} J. D. Barrow and A. C. Ottewill, J. Phys. A: Math. Gen. \textbf{16},
2757 (1983).

\bibitem{Bu5} H. Kleinert and H.-J. Schmidt, Gen. Rel. Grav. \textbf{34},
1295 (2002).

\bibitem{fR1} S. M. Carroll, V. Duvvuri, M. Trodden, and M. S.
Turner, Phys. Rev. D \textbf{70}, 043528 (2004).

\bibitem{fRn1} W. Hu and I. Sawicki, Phys. Rev. D {\bf 76}, 064004 (2007).

\bibitem{fRn2}  S. A. Appleby and R. A. Battye, Phys. Lett. B {\bf 654}, 7 (2007).

\bibitem{fRn3} A. A. Starobinsky, JETP Lett. {\bf 86}, 157 (2007).

\bibitem{fR2} V.~Faraoni, Phys.\ Rev.\ D \textbf{75}, 067302
(2007).

\bibitem{fR3} C.~G.~B\"ohmer, L.~Hollenstein and F.~S.~N.~Lobo, Phys.\ Rev.\ D
\textbf{76}, 084005 (2007).

\bibitem{fR4} C. G. Boehmer, T. Harko, and F. S. N. Lobo, Astropart. Phys. {\bf 29}, 386 (2008).

\bibitem{fR5}  C. S. J. Pun, Z. Kovacs, and T. Harko, Phys. Rev. \textbf{D 78}, 024043 (2008).

\bibitem{fR6} C.~G.~Boehmer, T.~Harko, and F.~S.~N.~Lobo, JCAP
\textbf{03}, 024 (2008).

\bibitem{fRn4}  S. A. Appleby, R. A. Battye and A. A. Starobinsky, JCAP {\bf 1006}, 005 (2010).

\bibitem{fR7} V. K. Oikonomou, F. P. Fronimos, and N. Th. Chatzarakis,
Physics of the Dark Universe {\bf 30},  100726 (2020).

\bibitem{fR8} A. V. Astashenok, S. Capozziello, S. D. Odintsov, and V. K. Oikonomou, Physics Letters {\bf B 816},  136222 (2021).

\bibitem{fR9} S. Chakraborty, K. MacDevette, and P. Dunsby, 	Phys. Rev. {\bf D 103}, 124040 (2021).

\bibitem{fR10} V. K. Oikonomou, Phys. Rev. {\bf D 103}, 044036 (2021).

\bibitem{fR11} M. A. Mitchell, C. Arnold, and B. Li, MNRAS {\bf 502}, 6101 (2021).

\bibitem{HMP1}  T. Harko, T. S. Koivisto, F. S. N. Lobo and G. J. Olmo,  Phys. Rev. {\bf D 85},  084016 (2012).

\bibitem{HMP2}  S. Capozziello, T. Harko, F. S. N. Lobo and G. J. Olmo, Int. J. Mod. Phys. {\bf D 22}, 1342006  (2013).
\bibitem{HMP3}  S. Capozziello, T. Harko, T. S. Koivisto, F. S. N. Lobo and G. J. Olmo,  JCAP {\bf 04},  011 (2013).
\bibitem{HMP4}  S. Capozziello, T. Harko, T. S. Koivisto, F. S. N. Lobo and G. J. Olmo,  Universe {\bf 1}, 199 (2015).

\bibitem{W1} Z. Haghani, T. Harko, H. R. Sepangi, and S. Shahidi, JCAP {\bf 10},  061 (2012).

\bibitem{W2} Z. Haghani, T. Harko, H. R. Sepangi, and S. Shahidi, Phys. Rev. {\bf D 88}, 044024 (2013).

\bibitem{W3}  J. M. Nester and H.-J. Yo, Chinese Journal of Physics
{\bf 37}, 113 (1999).

\bibitem{W4} J. Beltr\'{a}n Jim\'{e}nez, L. Heisenberg, and T. Koivisto,
Phys. Rev. D 98, 044048 (2018).

\bibitem{W5} R. D’Agostino and O. Luongo, Phys. Rev. {\bf D 98}, 124013
(2018).
\bibitem{W6}  M. Fontanini, E. Huguet, and M. Le Delliou, Phys. Rev.
{\bf D 99}, 064006 (2019).
\bibitem{W7} T. Koivisto and G. Tsimperis, Universe {\bf 5}, 80 (2019).

\bibitem{W8}  J. G. Pereira and Y. N. Obukhov, Universe {\bf 5}, 139
(2019).
 \bibitem{W9} D. Blixt, M. Hohmann, and C. Pfeifer, Universe {\bf 5}, 143
(2019).

\bibitem{W10} A. A. Coley, R. J. van den Hoogen, and D. D. McNutt,
Journal of Mathematical Physics {\bf 61}, 072503 (2020).
\bibitem{weylcartan} Z. Haghani, N. Khosravi and S. Shahidi, Class. Quant. Grav. {\bf 32}, 215016 (2015).
\bibitem{R1} T.~P.~Sotiriou and V.~Faraoni, Rev. Mod. Phys. \textbf{82},
451 (2010).

\bibitem{R2} A.~De Felice and S.~Tsujikawa, %``f(R) theories,''
Living Rev.\ Rel.\ \textbf{13}, 3 (2010).

\bibitem{R3} Y. F. Cai, S. Capozziello, M. De Laurentis and E. N.
Saridakis, Rept. Prog. Phys. {\bf 79}, 106901 (2016).

\bibitem{R4} S. Nojiri, S. D. Odintsov, and V. K.  Oikonomou, Physics Reports {\bf  692},  1 (2017).

\bibitem{book} T. Harko and F. S. N. Lobo, Extensions of f(R) Gravity:
Curvature-Matter Couplings and Hybrid Metric-Palatini Theory, Cambridge University Press, Cambridge, UK, 2018

\bibitem{R5} D. Langlois, Int. J. Mod. Phys. {\bf D 28} 1942006-3287 (2019).

\bibitem{fLm1} O.~Bertolami, C.~G.~Boehmer, T.~Harko, and F.~S.~N.~Lobo, Phys. Rev. \textbf{ D 75}, 104016 (2007).

\bibitem{fLm2} T.~Harko, Phys. Lett. \textbf{B 669}, 376 (2008).

\bibitem{fLm3} T. Harko and F. S. N. Lobo,  Eur. Phys. J. {\bf C 70}, 373 (2010).

\bibitem{fLm4} T. Harko, Phys. Rev. \textbf{D 81}, 044021 (2010).

\bibitem{fLm5} T. Harko and F. S. N. Lobo, Phys. Rev {\bf D 86}, 124034 (2012).

\bibitem{fLm6} J. Wang and K. Liao,  Class. Quant. Grav. {\bf 29},  215016 (2012).

\bibitem{fLm6a} O. Minazzoli and T. Harko, Phys. Rev. {\bf D 86}, 087502 (2012).

\bibitem{fLm6b} T. Harko, F. S. N. Lobo, and O. Minazzoli, Phys. Rev. {\bf D 87}, 047501 (2013).

\bibitem{fLm7} D. W. Tian and I. Booth, Phys. Rev. {\bf D 90},  024059 (2014).

\bibitem{fLm8} T. Harko, F. S. N. Lobo, J. P. Mimoso, and D. Pav\'{o}n, Eur. Phys. J. {\bf C 75},  386 (2015).

\bibitem{fLm9} R. P. L. Azevedo and P. P. Avelino, Phys. Rev. {\bf D 98}, 064045 (2018).

\bibitem{fLm10} S. Bahamonde, Eur. Phys. J. {\bf C  78}, 326 (2018).

\bibitem{fLm11} R. March, O. Bertolami, M. Muccino, R. Baptista, and S. Dell'Agnello, Phys. Rev. {\bf D 100}, 042002 (2019).

\bibitem{fLm12} O. Bertolami and C. Gomes, Phys. Rev. {\bf D 102}, 084051 (2020).

\bibitem{fT1}  T. Harko, F. S. N. Lobo, S. Nojiri and S. D. Odintsov,  Phys. Rev. {\bf D 84},  024020 (2011).

\bibitem{fT2} F. G. Alvarenga, A. de la Cruz-Dombriz, M. J. S. Houndjo, M.
E. Rodrigues, and D. Saez-Gomez, Phys. Rev. D \textbf{87}, 103526 (2013).

\bibitem{fT3} T. Harko, Phys. Rev. {\bf D 90}, 044067 (2014).

\bibitem{fT4} E. H. Baffou, M. J. S. Houndjo, M. E. Rodrigues, A. V.
Kpadonou, and J. Tossa, Phys. Rev. {\bf D 92}, 084043 (2015).

\bibitem{fT5} P. H. R. S. Moraes, J. D. V. Arbanil, and M. Malheiro, JCAP
\textbf{06}, 005 (2016).

\bibitem{fT6} M. E. S. Alves, P. H. R. S. Moraes, J. C. N. de Araujo, and
M. Malheiro, Phys. Rev. {\bf D 94}, 024032 (2016).

\bibitem{fT7} M. Zubair, S. Waheed, and Y. Ahmad, Eur. Phys. J. {\bf C 76%
}, 444 (2016).

\bibitem{fT8} M.-X. Xu, T. Harko, and S.-D. Liang, Eur. Phys. J. C \textbf{%
76}, 449 (2016).

\bibitem{fT9} H. Shabani and A. H. Ziaie, Eur. Phys. J. {\bf C 77}, 31
(2017).

\bibitem{fT10} P. H. R. S. Moraes and P. K. Sahoo, Phys. Rev. D \textbf{96}%
, 044038 (2017).

\bibitem{fT11} E. H. Baffou, M. J. S. Houndjo, M. Hamani-Daouda, and F. G.
Alvarenga, Eur. Phys. J. {\bf C 77}, 708 (2017).

\bibitem{fT12} F. Rajabi and K. Nozari, Phys. Rev. \textbf{D 96}, 084061
(2017).

\bibitem{fT12a} H. Velten and T. R. P. Carame\^{s}, Phys. Rev. {\bf D 95}, 123536 (2017).

\bibitem{fT13} P. K. Sahoo, P. H. R. S. Moraes, and P. Sahoo, Eur. Phys. J.
{\bf C 78}, 46 (2018).

\bibitem{fT13a} J. K. Singh, K. Bamba, R. Nagpal, and S. K. J. Pacif, Phys. Rev. {\bf D 97}, 123536  (2018).

\bibitem{fT14} J. Wu, G. Li, T. Harko, and S.-D. Liang,  Eur. Phys. J. {\bf C 78}, 430 (2018).

\bibitem{fT14a} R. Nagpal, S. K. J. Pacif, J. K. Singh, K. Bamba, and A. Beesham, Eur. Phys. J. {\bf C 78},  946 (2018).

\bibitem{fT14b} P. H. R. S. Moraes,  R. A. C. Correa, and G. Ribeiro, Eur. Phys. J. {\bf C  78},  192 (2018).

\bibitem{fT15} D. Deb, S. V. Ketov, M. Khlopov, and S. Ray, Journal of Cosmology and Astroparticle Physics {\bf 10},  070 (2019).

\bibitem{fT15a} P. H. R. S. Moraes, Eur. Phys. J. {\bf C 79}, 674  (2019).

\bibitem{fT16} S. K. Maurya, A. Errehymy, D. Deb, F. Tello-Ortiz, and M. Daoud, Phys. Rev. {\bf D 100}, 044014 (2019).

\bibitem{fT17} S. K. Maurya, A. Errehymy, K. Newton Singh, F. Tello-Ortiz, and M. Daoud, Physics of the Dark Universe {\bf 30},  100640 (2020).

\bibitem{fT18} R. Lobato, O. Lourenco, P. H. R. S. Moraes, C. H. Lenzi, M. de Avellar, W. de Paula, M. Dutra, and M. Malheiro, Journal of Cosmology and Astroparticle Physics {\bf 12},  039 (2020).

\bibitem{fT19} T. Harko and P. H. R. S. Moraes, Phys. Rev. {\bf D 101}, 108501 (2020).

\bibitem{fT20} G. A. Carvalho, F. Rocha, H. O. Oliveira, and R. V. Lobato, Eur. Phys. J. {\bf C  81}, 134 (2021).

\bibitem{fT21} M. Gamonal, Phys. Dark Univ. {\bf 31}, 100768 (2021).

\bibitem{fT22} J. M. Z. Pretel, S. E. Jor\'{a}s, R. R. R. Reis, and J. D. V. Arba$\tilde{{\rm n}}$il, 	JCAP {\bf 04},  064 (2021).

\bibitem{fT23} P. V. Tretyakov, Eur. Phys. J. {\ C 78},  896 (2018).

\bibitem{fTT} T. Harko, F. S. N. Lobo, G. Otalora, and E. N. Saridakis, JCAP {\bf 12},  021 (2014).

\bibitem{fQC1} Y. Xu, G. Li, T. Harko, and S.-D. Liang, Eur. Phys. J. {\bf C 79}, 708 (2019).

\bibitem{fQC2} Y. Xu, T. Harko, S. Shahidi, and S.-D. Liang, Eur. Phys. J. {\bf C 80}, 449 (2020).

\bibitem{fQC3} J.-Z. Yang, S. Shahidi, T. Harko, and S.-D. Liang, Eur. Phys. J. {\bf C 81}, 111 (2021).

 \bibitem{M1} R. Myrzakulov, \textit{Dark Energy in F(R,T) Gravity},  [arXiv:1205.5266]

\te{ \bibitem{M1a} D. Iosifidis, N. Myrzakulov, and R. Myrzakulov, Universe {\bf 7},  262 (2021).}

 \bibitem{M1b} N. Myrzakulov, R. Myrzakulov, and L. Ravera, eprint arXiv:2108.00957 (2021).

\bibitem{M1c} K. Yesmakhanova, N. Myrzakulov, S. Myrzakul, G. Yergaliyeva, K. Myrzakulov, K. Yerzhanov, and R. Myrzakulov,  [arXiv:2101.05318] (2021).

\bibitem{P-M} I. Prigogine, J. Geheniau, E. Gunzig, and P. Nardone,
Proceedings Of The National Academy Of Sciences \textbf{85}, 7428 (1988).

\bibitem{Lima} M. O. Calvao, J. A. S. Lima, and I. Waga, Phys. Lett. 
\textbf{A 162}, 223 (1992).

\bibitem{Su} J. Su, T. Harko, and S.-D. Liang, Advances in High Energy
Physics \textbf{2017}, 7650238 (2017).

\bibitem{Bar} J. A. S. Lima and I. P. Baranov, Phys. Rev.  \textbf{D 90},
043515 (2014).

\bibitem{hubble} H. Boumaza and K. Nouicer, Phys. Rev. {\bf D 100},  124047 (2019).

\bibitem{ZH} Z. Haghani and T. Harko, European Phys. J.  {\bf C 81}, 615 (2021).

\bibitem{M2} E. N. Saridakis, S.  Myrzakul, K. Myrzakulov, and K. Yerzhanov, Phys. Rev. {\bf D 102}, 023525 (2020).

\bibitem{M3} F. K. Anagnostopoulos,  S.  Basilakos, and E. N. Saridakis,  Phys. Rev. {\bf D 103}, 104013 (2021).

\bibitem{M5}  G. Bauyrzhan, N.  Myrzakulov, N. Serikbayev, and K.  Yerzhanov, \textit{Cosmology in Myrzakulov gravity-V}, [https://www.researchgate.net/publication/351022054. Cosmology in Myrzakulov gravity-V] (2021).
    
\bibitem{Toms} L. Parker and D. Toms, Quantum Field Theory in Curved Spacetime:
Quantized Fields and Gravity, Cambridge University Press, Cambridge, 2011    
    
\bibitem{Phys} Z. Haghani and T. Harko, Physics {\bf 3},  689 (2021).
    
\bibitem{sg3a} T. W. B. Kibble and S. Randjbar-Daemi, J. Phys. A:
Math. Gen. {\bf 13}, 141 (1980).
\end{thebibliography}
 \end{document}